\documentclass[sigconf,authorversion=true]{acmart}

\AtBeginDocument{%
  \providecommand\BibTeX{{%
    \normalfont B\kern-0.5em{\scshape i\kern-0.25em b}\kern-0.8em\TeX}}}

\copyrightyear{2021}
\acmYear{2021}
\setcopyright{acmlicensed}
\acmConference[MobiSys '21]{The 19th Annual International Conference on Mobile Systems, Applications, and Services}{June 24-July 2, 2021}{Virtual, WI, USA}
\acmBooktitle{The 19th Annual International Conference on Mobile Systems, Applications, \\ and Services (MobiSys '21), June 24-July 2, 2021, Virtual, WI, USA}
\acmPrice{15.00}
\acmDOI{10.1145/3458864.3467883}
\acmISBN{978-1-4503-8443-8/21/06}

\usepackage{booktabs} 
\usepackage[acronym, nowarn]{glossaries}
\makeglossaries

\newacronym{ai}{AI}{artificial intelligence}
\newacronym{ap}{AP}{access point}
\newacronym{ble}{BLE}{Bluetooth Low Energy}
\newacronym{iot}{\textsc{IoT}}{Internet of Things}
\newacronym{sdp}{\textsc{SDP}}{secure device paring}
\newacronym{zip}{\textsc{ZIP}}{zero-interaction pairing}
\newacronym{zia}{\textsc{ZIA}}{zero-interaction authentication}
\newacronym{zis}{\textsc{ZIS}}{zero-interaction security}
\newacronym{dc}{\textsc{DC}}{device context}
\newacronym{ac}{\textsc{AC}}{application context}
\newacronym{tar}{TAR}{True Acceptance Rate}
\newacronym{far}{FAR}{False Acceptance Rate}
\newacronym{frr}{FRR}{False Rejection Rate}
\newacronym{eer}{EER}{Equal Error Rate}
\newacronym{pki}{PKI}{Public Key Infrastructure}
\newacronym{fft}{FFT}{fast Fourier transform}
\newacronym{ntp}{NTP}{Network Time Protocol}
\newacronym{imu}{IMU}{inertial measurement unit}
\newacronym{csi}{CSI}{channel state information}
\newacronym{fpake}{fPAKE}{Fuzzy Password-Authenticated Key Exchange}
\newacronym{snr}{SNR}{signal-to-noise ratio}
\newacronym{sg}{SG}{Savitzky-Golay}
\newacronym{ewma}{EWMA}{exponentially weighted moving average}
\newacronym{vanet}{VANET}{vehicular ad-hoc network}
\newacronym{ecc}{ECC}{error correction code}
\newacronym{pake}{PAKE}{password-authenticated key exchange}
\newacronym{eke}{EKE}{Encrypted Key Exchange}
\newacronym{ecu}{ECU}{Electronic Control Unit}

\usepackage{amsmath,amsfonts,bbm}
\usepackage{algorithmic}
\usepackage{graphicx}
\usepackage{textcomp}
\usepackage{xcolor}
\usepackage{cryptocode}
\usepackage{caption}
\usepackage{subcaption}
\usepackage{xspace}
\usepackage{ulem}
\usepackage{tabularx}

\usepackage{hyperref}

\usepackage{enumitem}

\usepackage{xargs}   
\usepackage{xcolor}
\usepackage{colortbl}  

\newcommand{\name}{\textit{FastZIP}\xspace} 

\newenvironment{tightitem}{\begin{itemize}\setlength{\itemsep}{1pt}\setlength{\parskip}{0pt}\setlength{\parsep}{0pt}}{\end{itemize}}

\usepackage{soul}

\usepackage{makecell}

\usepackage{multirow}

\DeclarePairedDelimiter\ceil{\lceil}{\rceil}

\usepackage{filecontents}

\usepackage{pifont}

\usetikzlibrary{arrows.meta}
\usetikzlibrary{shapes.geometric}
\usetikzlibrary{backgrounds}
\usepackage{pgfplots}
\usepackage{capt-of}

\definecolor{mygreen}{HTML}{15b01a}
\definecolor{myred}{HTML}{e50000}

\begin{document}

\title{FastZIP: Faster and More Secure Zero-Interaction Pairing}

\author{Mikhail Fomichev}
\email{mfomichev@seemoo.tu-darmstadt.de}
\orcid{0000-0001-9697-0359}
\affiliation{%
  \country{Technical University of Darmstadt}
}
  
\author{Julia Hesse}
\email{jhs@zurich.ibm.com}
\affiliation{%
  \country{IBM Research Europe - Zurich}
}

\author{Lars Almon}
\email{lalmon@seemoo.tu-darmstadt.de}
\orcid{0000-0003-1296-2920}
\affiliation{%
  \country{Technical University of Darmstadt}
}

\author{Timm Lippert}
\email{timm.lippert@gmail.com}
\affiliation{%
  \country{Technical University of Darmstadt}
}

\author{Jun Han}
\email{junhan@comp.nus.edu.sg}
\affiliation{%
  \country{National University of Singapore}
}

\author{Matthias Hollick}
\email{mhollick@seemoo.tu-darmstadt.de}
\orcid{0000-0002-9163-5989}
\affiliation{%
  \country{Technical University of Darmstadt}
}

\renewcommand{\shortauthors}{M. Fomichev et al.}

\begin{abstract}
With the advent of the \gls{iot}, establishing a secure channel between smart devices becomes crucial.
Recent research proposes \textit{\gls{zip}}, which enables pairing without user assistance by utilizing devices' physical context (e.g., ambient audio) to obtain a shared secret key. 
The state-of-the-art \gls{zip} schemes suffer from three limitations: (1) prolonged pairing time (i.e., minutes or hours), (2) vulnerability to brute-force offline attacks on a shared key, and (3) susceptibility to attacks caused by predictable context (e.g., replay attack) because they rely on limited entropy of physical context to protect a shared key.
We address these limitations, proposing \name, a novel \gls{zip} scheme that significantly reduces pairing time while preventing offline and predictable context attacks.  
In particular, we adapt a recently introduced \textit{\gls{fpake}} protocol and utilize \textit{sensor fusion}, maximizing their advantages. 
We instantiate \name for intra-car device pairing to demonstrate its feasibility and show how the design of \name can be adapted to other \gls{zip} use cases. 
We implement \name and evaluate it by driving four cars for a total of 800~km. We achieve up to \textit{three times shorter} pairing time compared to the state-of-the-art \gls{zip} schemes while assuring robust security with adversarial \textit{error rates below 0.5\%}.

\end{abstract}


%

\maketitle


\glsresetall

\section{Introduction}
\label{sec:intro}
The proliferation of the \gls{iot} urges the need to secure wireless communication between smart devices to protect data they exchange (e.g., sensor readings). 
Such protection is crucial to ensure user privacy and trustworthiness of \gls{iot} systems~\cite{Han:2018, Lin:2019, Lee:2019}. 
To secure wireless communication, unassociated devices need to establish a shared secret key---a process known as \textit{secure pairing}. 
This shared key is used by devices to provide encryption and authentication. 
In recent years, numerous pairing schemes have been proposed, most of which rely on user assistance (e.g., entering a password)~\cite{Mirzadeh:2014, Chong:2014, Fomichev:2017}.
However, many \gls{iot} devices are not equipped with user interfaces, making user-assisted pairing impractical~\cite{Han:2018, Lee:2019}. 
In addition, a rapid increase in the number of smart devices limits scalability of user-assisted schemes~\cite{Fomichev:2019perils, Schurmann:2017}. 
 
To address this problem, recent research proposes \textit{\gls{zip}} utilizing devices' context to derive a shared secret key without user involvement~\cite{Schurmann:2013, Miettinen:2014, Han:2018, Fomichev:2019}. Such context is represented as a set of \textit{sensor modalities} (e.g., audio, acceleration) collected by devices from their ambient environment.
\gls{zip} schemes utilize \textit{colocated} devices residing in an enclosed physical space such as a car to observe similar context compared to devices outside. 
Specifically, the colocated devices record their context and translate it to sequences of bits called  \textit{fingerprints}, which are input to a key agreement protocol to establish a shared cryptographic key.
Thus, the security of \gls{zip} schemes relies on the unpredictability of context, which depends on the intensity and variety of ambient activity (e.g., sound, motion) occurring in the environment. 

To date, a number of \gls{zip} schemes utilizing various sensor modalities to capture context have been proposed~\cite{Schurmann:2013, Miettinen:2014, Schurmann:2017, Han:2017, Han:2018, Miettinen:2018, Lin:2019}. 
These state-of-the-art schemes have three major limitations: (1) prolonged pairing time, (2) vulnerability to offline attacks, and (3) susceptibility to attacks caused by predictable context (e.g., replay). 
First, state-of-the-art \gls{zip} schemes suffer from \textit{prolonged pairing time} requiring minutes and hours of context data to establish a shared key~\cite{Han:2018, Fomichev:2019perils}. 
This happens because they use a cryptographic primitive called \textit{fuzzy commitments}~\cite{fuzzyCom}, where the entropy of a shared key \textit{is equal to} the entropy of fingerprint bits, input to the protocol. 
Thus, these \gls{zip} schemes need to obtain at least 128 bits of entropy from context to ensure that a shared key provides adequate security~\cite{Bluekrypt:2020}. 
Obtaining these bits takes a prolonged time because many contexts change slowly.
Second, state-of-the-art \gls{zip} schemes are by design vulnerable to \textit{offline attacks}, namely an adversary can mount a brute-force attack on a shared key by repeatably guessing the used fingerprints. 
These schemes can \textit{only} withstand offline attacks if they use (1) long fingerprints (i.e., \textgreater 128 bits) of (2) high entropy. 
However, recent works find severe entropy biases (e.g., bit patterns) in fingerprints of state-of-the-art \gls{zip} schemes~\cite{Bruesch:2019, Fomichev:2019perils}, exposing them to offline attacks.
Third, state-of-the-art \gls{zip} schemes are susceptible to context replay, inference, or monitoring attacks due to \textit{predictable context}~\cite{Han:2017, Fomichev:2019perils, Bruesch:2019}. 
Frequently, context becomes predictable because it relies on a single sensor modality (e.g., acceleration). 
Thus, an adversary can obtain similar context in comparable environments, use better hardware, or employ video analysis. 

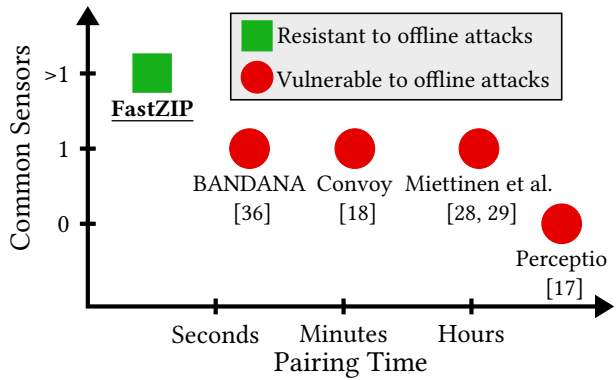
\begin{figure}
\centering
	\begin{tikzpicture}[square/.style={regular polygon,regular polygon sides=4}]
		\draw [>=Triangle, <->, ultra thick] (0,4) node (yaxis) {} |- (7,0) node (xaxis) {};
		
		\draw [ultra thick] (3pt,1.1) -- (-3pt,1.1) node[anchor=east] {\large 0}; 
		\draw [ultra thick] (3pt,2.1) -- (-3pt,2.1) node[anchor=east] {\large 1};
		\draw [ultra thick] (3pt,3.1) -- (-3pt,3.1) node[anchor=east] {\large \textgreater1};
		
		\draw [ultra thick] (1.7,3pt) -- (1.7,-3pt) node[anchor=north] {\large Seconds};
		\draw [ultra thick] (3.4,3pt) -- (3.4,-3pt) node[anchor=north] {\large Minutes};
		\draw [ultra thick] (5.1,3pt) -- (5.1,-3pt) node[anchor=north] {\large Hours};
		
		\draw [mygreen, fill=mygreen, text=black] (0.6, 2.85) rectangle ++(0.51cm, 0.51cm) 
		node[xshift=-0.25cm, yshift=-0.5cm, anchor=north] {\normalsize \underline{\textbf{FastZIP}}}; 
		
		\draw[myred,fill=myred, text=black] (2.15, 2.1) circle (7.5pt) node[yshift=-0.18cm, anchor=north] 
		{\normalsize \begin{tabular}{c} BANDANA \\~\cite{Schurmann:2017}\end{tabular}}; 
		
		\draw[myred,fill=myred, text=black] (3.55, 2.1) circle (7.5pt) node[yshift=-0.18cm, anchor=north] 
		{\normalsize \begin{tabular}{c} Convoy \\~\cite{Han:2017}\end{tabular}}; 
		
		\draw[myred,fill=myred, text=black] (5.2, 2.1) circle (7.5pt) node[yshift=-0.18cm, anchor=north] 
		{\normalsize \begin{tabular}{c} Miettinen et al. \\~\cite{Miettinen:2014, Miettinen:2018}\end{tabular}}; 
		
		\draw[myred,fill=myred, text=black] (6.3, 1.1) circle (7.5 pt) node[yshift=-0.18cm, anchor=north] 
		{\normalsize \begin{tabular}{c} Perceptio \\~\cite{Han:2018}\end{tabular}}; 
		
		\draw [black, semithick, fill=gray!15, show background rectangle] (1.9, 3.9) rectangle ++(4.4, -1.15);
		
		\draw [mygreen, fill=mygreen, text=black] (2.05, 3.4) rectangle ++(0.4cm, 0.4cm) 
		node[xshift=-0.04cm, yshift=-0.19cm, anchor=west] {\normalsize Resistant to offline attacks};
		
		\draw[myred,fill=myred, text=black] (2.25, 3.05) circle (6pt)
		node[xshift=0.15cm, yshift=-0.02cm, anchor=west] {\normalsize Vulnerable to offline attacks};
		
		\node[xshift=-3.5cm, below=0.5cm] at (xaxis) {\Large Pairing Time};
		\node[yshift=-0.4cm, rotate=90, left=0.9cm] at (yaxis) {\Large Common Sensors};
	\end{tikzpicture}
	\caption{Design space of \name: it provides shorter pairing time and improved security compared to state-of-the-art \gls{zip} schemes utilizing more common sensors.}
	\label{fig:fz-design}
\end{figure}

The above three limitations impair practicality and security of \gls{zip} schemes, hindering their real-world deployment. 
To overcome these limitations, we propose \name, a novel \gls{zip} scheme that achieves shorter pairing time and improved security by addressing the following two challenges. 
~\autoref{fig:fz-design} compares \name and state-of-the-art \gls{zip} schemes in terms of paring time\footnote{We use the shortest pairing time reported in the original publication for each scheme.}, resistance to offline attacks, and the number of common sensors required for pairing.  

First, to shorten pairing time, we need to reduce the number of fingerprint bits while being robust to offline attacks. 
This is challenging because fewer bits means less entropy in a shared key, easing an offline attack. 
To address this challenge, we adapt a recently introduced \textit{\gls{fpake}} protocol~\cite{Fpake:2018}. 
\gls{fpake} establishes a shared key from low-entropy secrets (e.g., short passwords) and is resistant to offline attacks.
While \gls{fpake} is an existing protocol, adapting it to \gls{zip} schemes is not trivial. 
Specifically, we find that \gls{fpake} protection against offline attacks is not always guaranteed in realistic \gls{zip} settings, namely when colocated devices do not yield highly similar fingerprints from context. 
Thus, we analyze how to set \gls{fpake} parameters to withstand offline attacks even in such settings (cf.~\autoref{sec:sysdesign}).  
To the best of our knowledge, we are the first to implement \gls{fpake} and demonstrate that it shortens pairing time and improves security of \gls{zip} schemes using real-world data. 

Second, to defend against predictable context attacks (e.g., replay attacks), we propose a simple form of \textit{sensor fusion} by concatenating fingerprints derived from different sensors, each capturing distinct ambient activity.
Applying sensor fusion is not straightforward, as we require a generic method to extract fingerprint bits of sufficient entropy from heterogeneous sensor signals (cf.~\autoref{sec:sysdesign}). 
Existing methods rely on scenario-specific characteristics of sensor signals (e.g., peak occurence), thus cannot be directly reused, and they often produce fingerprints with entropy biases~\cite{Fomichev:2019perils, Bruesch:2019}.
We demonstrate that sensor fusion not only prevents predictable context attacks (e.g., replay) but also assists \gls{fpake} in shortening paring time, as we obtain more bits from context, accumulating entropy faster. 
Sensor fusion is feasible because smart devices have multiple sensor modalities often \textit{integrated} in one chip, for example, an \gls{imu} contains an accelerometer, gyroscope, and magnetometer~\cite{IMU:2019}, while a camera has light and RGB sensors, and a wireless chipset hosts both Wi-Fi and Bluetooth~\cite{Ruge:2020}.

We demonstrate the advantages of \name by evaluating a novel use case of \textit{intra-car device pairing} (cf.~\autoref{sec:eval}), which is inspired by the growing number of smart devices inside modern cars. 
For example, the increasing popularity of carsharing and self-driving rides urges the need to pair multiple user devices (e.g., smartphone, earbuds) with infotainment systems of different cars to enable such services as customized driving experience~\cite{Sanz:2016, Hertz:2020}. Furthermore, \glspl{ecu} require pairing with wireless third-party components (e.g., tire pressure monitor) to enable travel efficiency and safety~\cite{Cho:2018, Claburn:2018, Trikutam:2019}. In both examples, the growing number of devices hinder manual pairing, requiring pairing solutions without user intervention.  
Despite focusing on in-car pairing, we show how the design of \name can generalize to other \gls{zip} use cases (e.g., smart home) to improve pairing time and security in~\autoref{sec:discuss}.

Through our real-world experiments, we demonstrate the feasibility of leveraging the context of a moving car to pair devices inside it. 
Such context is affected by road and traffic conditions, car characteristics such as suspension, and driving patterns, and it can be captured by accelerometer, gyroscope, and barometer sensors~\cite{Sankaran:2014, Han:2017, Chen:2017, Vaas:2018} that are ubiquitous in user devices (e.g., smartphone) and modern cars. 
We evaluate \name by collecting  sensor data from four cars driven over 800~km on different road types, including urban, rural, and highways.  
In our evaluation, we assume that pairing devices can start measuring context simultaneously by receiving a broadcast command from the car's infotainment system, which is not compromised.
\name achieves up to \textit{three times} faster pairing compared to state-of-the-art \gls{zip} schemes, shows error rates \textit{below 0.5\%} in the presence of a powerful adversary, and runs efficiently on off-the-shelf \gls{iot} devices.
In summary, we make the following contributions:
\begin{tightitem}
	\item We design \name, a novel \gls{zip} scheme utilizing \gls{fpake} and sensor fusion to reduce pairing time and improve security.
	\item{We implement \name for intra-car device pairing and evaluate it by collecting real-world driving data, demonstrating the effectiveness of \name.}
	\item We publicly release the collected data, source code of our evaluation stack, and the first implementation of \gls{fpake}.
\end{tightitem}


\section{Background}
\label{sec:bkgrd}
\begin{figure}
\centering   
\newcommand{\mnode}[1]{\node(0,0) (#1) {};}

\definecolor{darkred}{RGB}{139,0,0}

\begin{tikzpicture}[every node/.style={font=\rmfamily\small, align=center, minimum height=0.5cm, minimum width=0.1cm}]

	\node [matrix, very thin,column sep=0.91cm,row sep=0.15cm] (matrix) at (0,0) {
	\mnode{A 0}	&             &             &             & \mnode{B 0} \\
	\mnode{A 1}	&             & \mnode{C 1}	&				      & \mnode{B 1} \\
	\mnode{A 2}	& \mnode{L 2}	& \mnode{C 2}	& \mnode{R 2}	& \mnode{B 2} \\
	\mnode{A 3}	& \mnode{L 3}	& \mnode{C 3} & \mnode{R 3}	& \mnode{B 3} \\
	\mnode{A 4}	&				      & \mnode{C 4}	&             & \mnode{B 4} \\
	\mnode{A 5}	&  				    & \mnode{C 5}	&             & \mnode{B 5} \\
	\mnode{A 6}	& 				    & \mnode{C 6}	&             & \mnode{B 6} \\
	\mnode{A 7}	& 				    & \mnode{C 7}	&             & \mnode{B 7} \\
	\mnode{A 8}	& 				    & \mnode{C 8} &             & \mnode{B 8} \\
};

\fill 
	(A 0) node[above] {\large \textbf{Device A}}
	(B 0) node[above] {\large \textbf{Device B}};

\draw [>=latex,line width=1pt] (A 0) -- (A 8);
\draw [>=latex,line width=1pt] (B 0) -- (B 8);

\draw [densely dashed,color=darkred,line width=1.5pt] 
  (A 1) -- (B 1) 
  (A 3.south east) -- (B 3.south west) 
  (A 5.south east) -- (B 5.south west);

\draw [-latex,line width=0.8pt] (A 2.north east) -- (L 2.north west); 
\draw [-latex,line width=0.8pt] (L 2.south west) -- (A 2.south east);
\draw [-latex,line width=0.8pt] (B 2.north west) -- (R 2.north east); 
\draw [-latex,line width=0.8pt] (R 2.south east) -- (B 2.south west);

\draw [-latex,line width=0.8pt] (A 4.south east) -- (B 4.south west);

\draw [-latex,line width=0.8pt] (A 6.south east) -- (B 6.south west);
\draw [latex-,line width=0.8pt] (A 7.south east) -- (B 7.south west);

\fill
 (C 1)        node[font=\normalsize, above] {\textcolor{darkred} {Entropy Amplification Phase}}
 (C 1.south)  node[font=\normalsize, below=1.5mm]					{Run PAKE $|f|-$times}
 (C 3.south)  node[font=\normalsize, above] {\textcolor{darkred}	{Fuzzy Commitment Phase}}
 (C 4.south)  node[font=\normalsize, above]             {$com = c \oplus \mathbf{k}$}
 (C 5.south)  node[font=\normalsize, above=] {\textcolor{darkred}	{Key Confirmation Phase}}
 (C 6.south)  node[font=\normalsize, above]             {$h$}
 (C 7.south)  node[font=\normalsize, above]	            {$h'$}

 
 (A 1.south)  node[font=\normalsize, below=2mm, left]   {Bits of $f$}
 (A 2.south)  node[font=\normalsize, above=1mm, left]	  {Vector $\mathbf{k}$}
 
 (B 1.south)  node[font=\normalsize, below=2mm, right]  {Bits of $f'$}
 (B 2.south)  node[font=\normalsize, above=1mm, right]	{Vector $\mathbf{k'}$}
 
 (A 4)        node[font=\normalsize, left]				      {Secret $s$}
 (A 4.south)	node[font=\normalsize, below=1mm, left]   {$c = \mathsf{ECC}(s)$}
 
 (B 4)        node[font=\normalsize, right]             {$c' = com\oplus \mathbf{k}'$}
 (B 4.south)  node[font=\normalsize, below=1mm, right]  {$s' = \mathsf{ECC}(c')$}
 
 (A 6)        node[font=\normalsize, left]              {$h = H(s||0)$}
 (A 7)        node[font=\normalsize, left]              {check $h'$}
 (A 8.north)  node[font=\normalsize, below=0mm, left]   {$k_{AB} = KDF(s)$}
 
 (B 6)        node[font=\normalsize, right]             {check $h$}
 (B 7)        node[font=\normalsize, right]             {$h' = H(s'||1)$}
 (B 8.north)  node[font=\normalsize, below=0mm, right]  {$k_{AB} = KDF(s')$};
 
 
\end{tikzpicture} 
\caption{Detailed flow diagram of the \gls{fpake} protocol.}
 \label{fig:fpakeschem-1}
\end{figure}
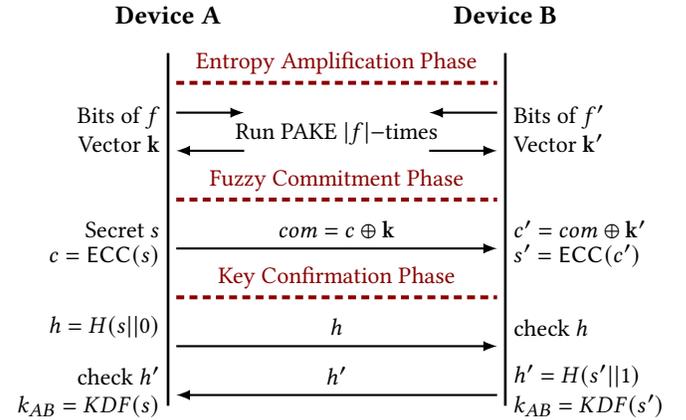

We first explain the working principle and shortcomings of fuzzy commitments---a cryptographic protocol used by state-of-the-art \gls{zip} schemes to share a secret key.
Then, we detail the \gls{fpake} protocol~\cite{Fpake:2018}, addressing these shortcomings, that we utilize in \name.  
\\
\textbf{\gls{zip} Based on Fuzzy Commitments.}
Prior work on \gls{zip} relies on fuzzy commitments or vaults~\cite{fuzzyCom,fuzzyVault} to exchange a key $K$ between two devices holding similar fingerprints $f,f'$~\cite{Schurmann:2013, Miettinen:2014, Schurmann:2017, Han:2017, Han:2018, Miettinen:2018, Lin:2019}.
Specifically, Device A chooses a 128-bit key $K$ and sends a commitment $c\gets\mathsf{ECC.Encode}(K)\oplus f$ to Device B, which can recover $K\gets\mathsf{ECC.Decode(c \oplus f')}$ if the fingerprint mismatch $f'\oplus f$ is within the error correction capability of the \gls{ecc}. 
While conceptually simple, this approach has two disadvantages in the case of \gls{zip}.
First, it inherently requires fingerprints $f,f'$ to be at least 140 bits, since they are XORed to an expanded encoding of the 128-bit key\footnote{The 140 bits are for an expanded encoding allowing up to 10\% mismatching fingerprint bits. To allow for 30\% mismatch in fingerprints, 205 bits are required.}.
Second, an eavesdropping adversary can capture the commitment $c$ and try decoding it with arbitrarily many fingerprint guesses to obtain the key $K$. 
This constitutes an \textit{offline attack} on $K$, which can only be defended against if the fingerprints have high entropy (i.e., they are hard to guess). 
In practice, state-of-the-art \gls{zip} schemes already require multiple minutes or even hours to obtain fingerprints \textgreater 128 bits from context~\cite{Schurmann:2017, Miettinen:2018, Han:2018, Han:2017, Miettinen:2014}. 
Even worse, an in-depth entropy analysis reveals that fingerprints of these schemes contain bit patterns or predictable distributions of 0- and 1-bits~\cite{Fomichev:2019perils, Bruesch:2019}. 
Thus, an adversary can more easily guess the fingerprints, exposing state-of-the-art schemes to offline attacks. 
\\
\textbf{fPAKE Protocol.}
\gls{fpake} used by \name allows reducing the number of required fingerprint bits, hence shortening pairing time, while providing resilience to offline attacks.  
In essence, \gls{fpake} is also a fuzzy commitment, but instead of creating the commitment from fingerprint $f$, \gls{fpake} adds an interactive \textit{entropy amplification} phase that turns fingerprints $f,f'$ into high entropy keys $\mathbb{k,k'}$ with a similar mismatch pattern as $f,f'$ (cf.~\autoref{fig:fpakeschem-1}).
In entropy amplification, \gls{fpake} leverages an established cryptographic primitive called password-authenticated key exchange (PAKE)~\cite{Bellovin:1992}, which allows two parties to exchange a secure (i.e., 128-bit and uniform) key from a shared short string, such as a password, or even a bit. 
The PAKE protocol is secure against offline attacks, meaning that the best possible adversarial strategy is to guess the short string and engage in the key exchange. 
In \gls{fpake}, PAKE is used to amplify the entropy of individual fingerprint bits as follows: Devices A and B run multiple standard PAKE~\cite{Bellovin:1992} protocols on the individual fingerprint bits in parallel, obtaining key vectors $\mathbb{k}$ and $\mathbb{k'}$, where $\mathbb{k}_i=\mathbb{k'}_i$ if the $i$-th fingerprint bits matched.
Next, Device A chooses a 128-bit secret $s$ and sends a fuzzy commitment $com\gets\mathsf{ECC.Encode}(s)\oplus \mathbb{k}$ to Device B, which decodes it with $\mathbb{k'}$. 
Afterwards, Devices A and B confirm to each other that they know $s$ by sending each other hash values $H(s||0)$ and $H(s'||1)$ of the secret. 
Finally, if the hash check succeeds, Devices A and B derive a shared key $k_{AB}$ from $s$ using a key derivation function (KDF). 
\\
\textbf{Advantages of fPAKE in \gls{zip}.}
By using high entropy keys in the fuzzy commitment phase (cf.~\autoref{fig:fpakeschem-1}), \gls{fpake} prevents an eavesdropping adversary from mounting an offline attack because the adversary only knows $c \oplus\mathbb{k}$, which is a secure encryption of $c$ under a (by the guarantees of PAKE) secure key $\mathbb{k}$.
Moreover, even an active adversary (e.g., malicious Device B) can try \textit{exactly one} fingerprint guess $f'$ as input to the interactive entropy amplification phase. 
If that one guess is too far (i.e., $f$ and $f'$ are dissimilar), even the unbounded adversary cannot recover $s$, making the offline attack \textit{impossible}. 
Otherwise, if the guess is ``close enough'' (cf.~\autoref{sec:sysdesign}), the active adversary can attempt an offline attack.
However, Device A waits for the key confirmation $h'$ within a short timeout (e.g., a few seconds), allowing the adversary only this amount of time to perform the attack. 
We note that for standard fuzzy commitments the key confirmation upon timeout cannot similarly limit the offline attack, as the adversary does not participate in the protocol. 

In~\autoref{sec:sysdesign}, we demonstrate how to leverage the strong security of \gls{fpake} against offline attacks to reduce the required fingerprint sizes \textit{well below} 128 bits in many settings.
Also, we empirically show that additional communication overhead of \gls{fpake} (i.e., entropy amplification phase) is negligible compared to up to three times faster pairing time when using \gls{fpake} instead of fuzzy commitments. 

\section{System and Threat Models}
\label{sec:mod}
We introduce our system model, describing the goal, requirements, and assumptions of \name, and
our threat model, detailing adversary's goals and capabilities. 
\\
\textbf{System Model.} 
\label{sub:sys-model}
The main \textit{goal} of \name is to establish a shared secret key between colocated devices within a trusted boundary (e.g., inside a car) based on the perceived context. 
We design \name to fulfill the following \textit{requirements}: (1) be free of user interaction during pairing (\textit{usability}), (2) have short pairing time (\textit{practicality}), and (3) work on commodity devices equipped with off-the-shelf sensors (\textit{deployability}). 
To achieve the main goal while satisfying the requirements, we make the following \textit{assumptions}: (1) devices running \name do not have any pre-shared secrets, nor any jointly trusted third party, (2) they communicate  using Wi-Fi or Bluetooth, and share a common set of sensors such as an accelerometer, gyroscope, and barometer, and (3) they begin measuring context upon receiving the ``start'' command from the car's infotainment system, which is assumed to be non-compromised.
\\
\textbf{Threat Model.} 
\label{sub:adv-model}
We consider an adversary whose \textit{goal} is to establish a shared secret key with a legitimate device while residing outside the trusted boundary.
In particular, the adversary attempts to either \textit{impersonate} one of the legitimate devices or acts as a \textit{man-in-the-middle} between a pair of devices. 
The adversary can neither compromise legitimate devices nor break cryptographic primitives, however, they fully control a wireless channel, are equipped with the same sensing hardware as legitimate devices, and have four attack capabilities. 
In an \textit{injection attack}, the adversary attempts to pair with legitimate devices using self-chosen context readings. In a \textit{replay attack}, the adversary replays precollected context readings. In a \textit{similar-context attack}, the adversary tries to actively match their context with legitimate devices. 
In the intra-car pairing, the adversary launching a replay attack replays the precollected context data from a route driven by a victim car carrying legitimate devices, while in a similar-context attack, they actively follow the victim car to capture similar context such as the road bumpiness. 
The first three attacks require the adversary to participate in the pairing protocol, while in an \textit{offline attack}, they record a successful pairing session and try to compute a shared key from it by repeatedly guessing fingerprints used by legitimate devices. 

\setlength{\belowdisplayskip}{1pt} \setlength{\belowdisplayshortskip}{1pt}
\setlength{\abovedisplayskip}{1pt} \setlength{\abovedisplayshortskip}{1pt}

\section{System Design}
\label{sec:sysdesign}
\begin{figure}
	\begin{center}
		\includegraphics[width=0.96\linewidth]{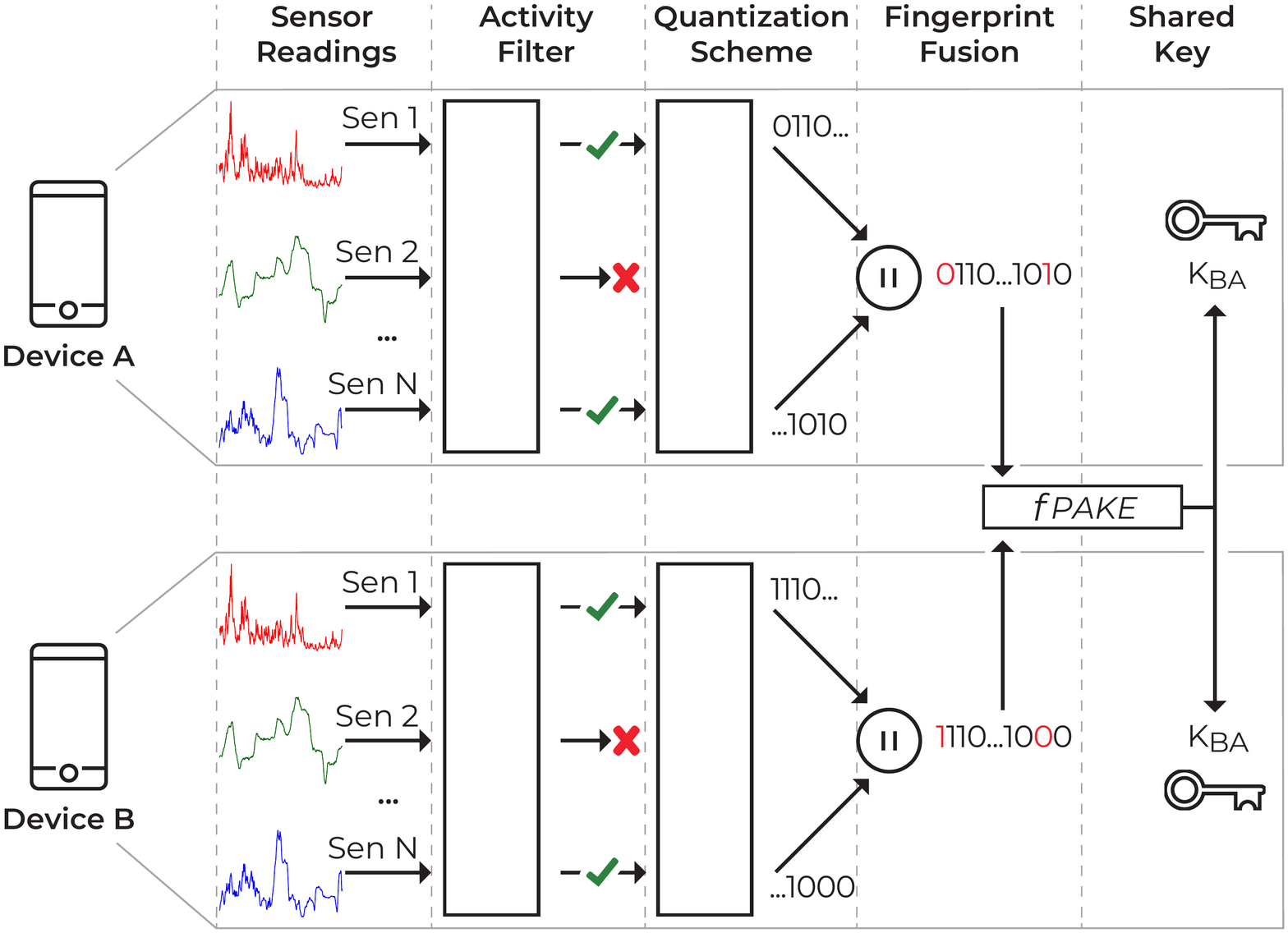}
	\end{center}
	\caption{System overview. \name takes as input a set of sensor readings from two devices. The readings are quantized to \textit{similar} fingerprints and afterwards input to the \gls{fpake} protocol to share a symmetric key, $K_{BA}$.}
	\label{fig:overview}
\end{figure}

We present the architecture of \name, describing its modules:  \textit{activity filter}, \textit{quantization}, and \textit{key exchange}. 
\\
\textbf{System Overview.}
The main goal of \name is to share a symmetric key between a pair of devices utilizing their context. 
In a moving car the context encompasses road turns, bumpiness, and speed changes~\cite{Vaas:2018, Sankaran:2014, Chen:2017}, and it can be perceived by accelerometer, gyroscope, and barometer sensors that are ubiquitous in smart devices.
\name works as follows (cf.~\autoref{fig:overview}): Devices \textit{A} and \textit{B} capture their context using a set of common sensors. 
The resulting sensor readings are input to the \textit{activity filter} to discard low-entropy context, which can be predicted by an adversary. 
Afterwards, the filtered sensor readings are input to the \textit{quantization} translating them into a sequence of  fingerprint bits. 
Each device constructs its fingerprint by concatenating sub-fingerprints derived from different sensors (i.e., sensor fusion). 
These fingerprints are input to the \textit{\gls{fpake}} protocol, which outputs a shared symmetric key if the fingerprints have a sufficient number of similar bits.
\begin{figure*}[t]
	\centering
	\begin{subfigure}[b]{.27\textwidth}
		\centering
		\includegraphics[width=\linewidth]{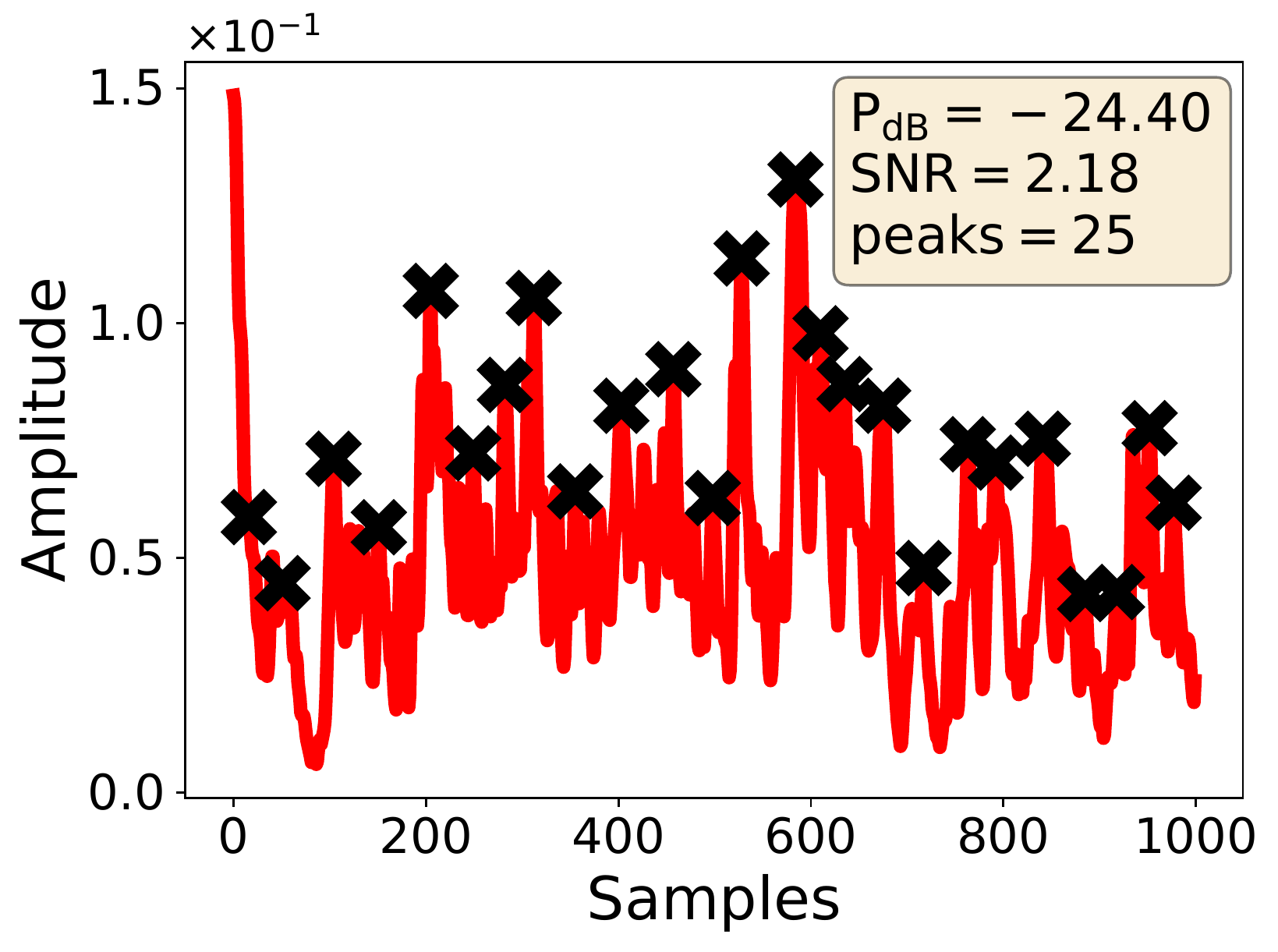}
		\caption{Signal with sufficient entropy}
		\label{sf:af-good}
  	\end{subfigure}
  	\begin{subfigure}[b]{.27\textwidth}
		\centering
		\includegraphics[width=\linewidth]{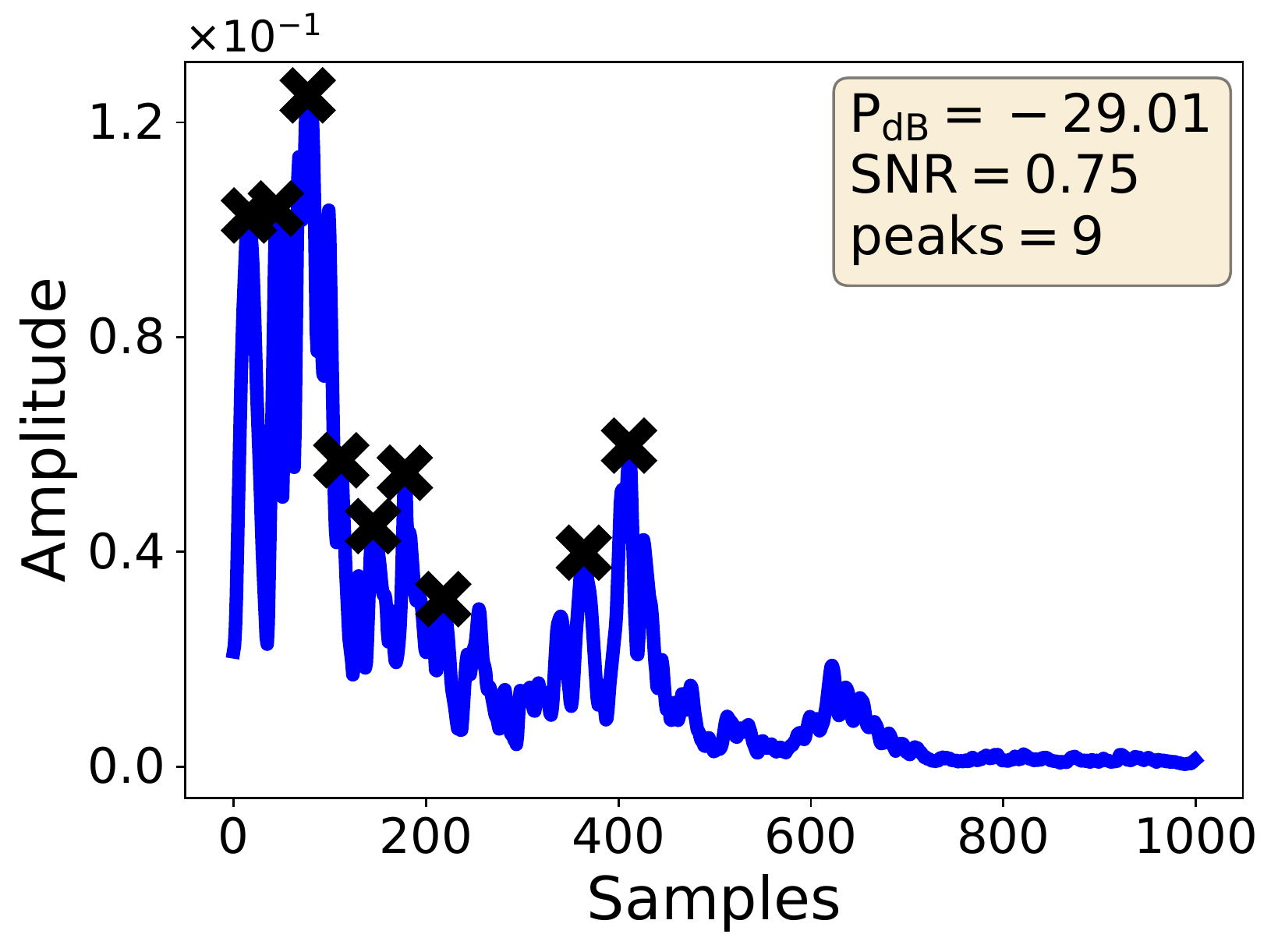} 
		\caption{Signal with insufficient entropy}
		\label{sf:af-bad}
  	\end{subfigure}
	\begin{subfigure}[b]{.27\textwidth}
		\centering
	    \includegraphics[width=\linewidth]{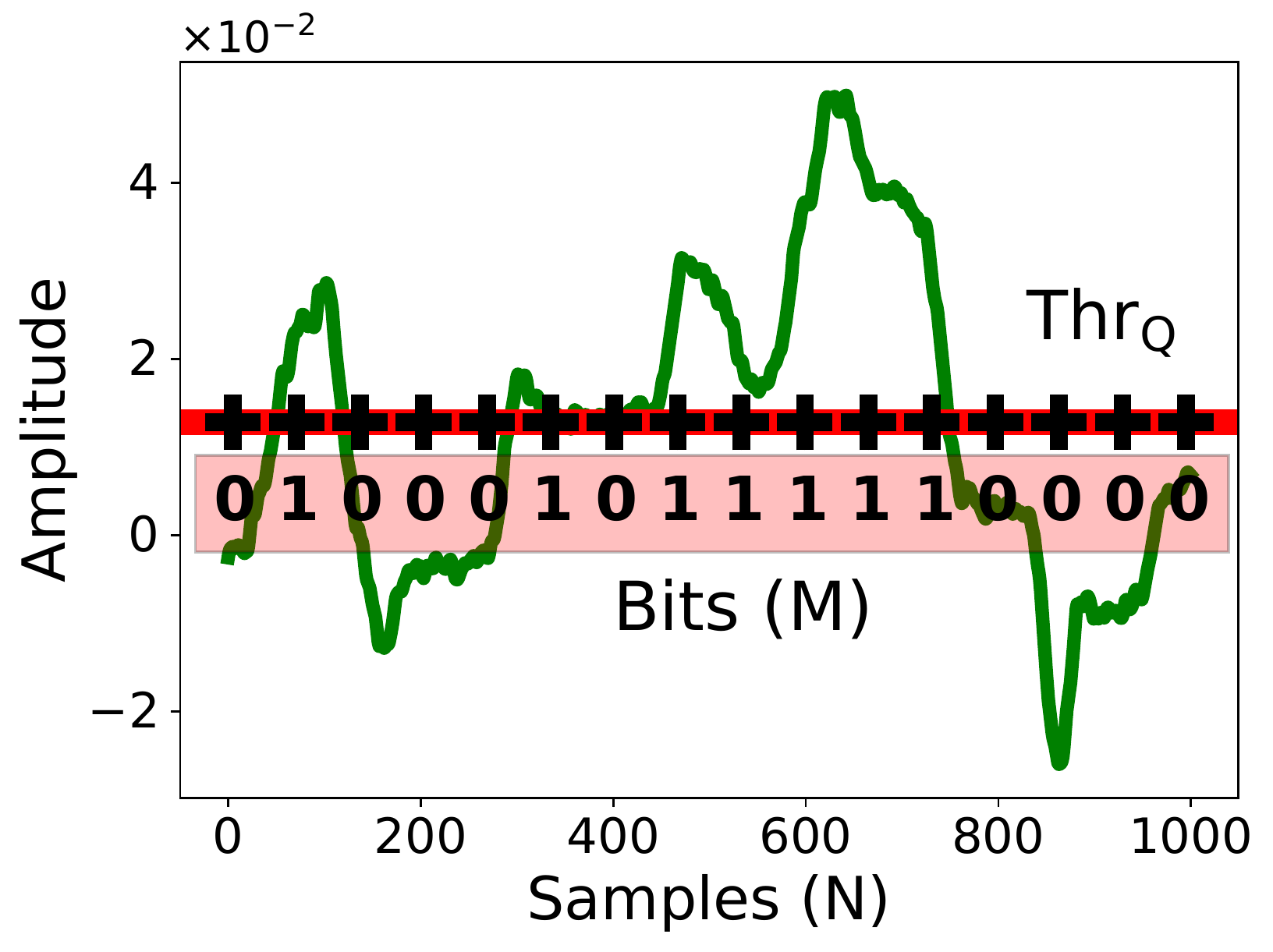}
	    \caption{Signal quantized to bits}
	    \label{sf:qs}
	\end{subfigure}
	\caption{Activity filter applied to acceleration signals (a) and (b); Quantization applied to a gyroscope signal (c).}
	\label{fig:ar-qs}
\end{figure*}
\\
\textbf{Activity Filter.}
\label{par:sysdes-ar}
The security of any \gls{zip} scheme relies on the unpredictability of context from outside a trusted boundary (e.g., car interior). The low-entropy context undermines security of \gls{zip} schemes, allowing an adversary to guess fingerprints derived from it~\cite{Fomichev:2019perils}. \name utilizes the \textit{activity filter} to ensure that fingerprints are obtained from context data with sufficient entropy.   

To estimate the entropy of a sensor signal, we analyze its strength relative to noise and variation.
For that, we employ three metrics: \textit{average power}, \textit{\gls{snr}}, and the \textit{number of prominent peaks}, which are used to characterize signal's quality~\cite{Karapanos:2015, Lin:2019, Vaas:2018}. The average power and \gls{snr} are applicable to all sensors, while prominent peaks is a complementary metric for rapidly changing modalities (e.g., acceleration), ensuring their sufficient variation. 
We compute the average power $P_s$ in dB of a discrete sensor signal $s(t)$ as follows: 
\[ P_{s (dB)} = 10\cdot\log_{10}\biggl(\dfrac{1}{T}\sum_{t=1}^{T}s^2(t)\biggr) \]
We cannot compute \gls{snr} as the ratio of signal to noise power, as we do not have the estimate of the latter; estimating noise power will impose additional processing overhead. 
Thus, we use an alternative definition of \gls{snr} as the ratio of mean to standard deviation of a signal: $SNR = \dfrac{\mu}{\sigma}$~\cite{Bushberg:2011}.
To find prominent peaks in a signal, we count peaks that have sufficient height relative to the highest peak, while being within minimum distance $\Delta_{P}$ from each other. Figures \ref{sf:af-good} and \ref{sf:af-bad} show activity filter metrics computed for two acceleration signals (prominent peaks marked with \ding{54}). 
We see that the former signal captures continuous activity, exhibiting sufficient entropy, while the latter signal contains noise in its right half, which is reflected in the computed metrics. 
 
After computing the metrics, we check them against fixed thresholds to discard signals with insufficient entropy, which, in turn, may reduce availability of \name. 
To avoid this, we apply the activity filter on a continuous stream of sensor data using an overlapping sliding window. Thus, parts of the signal containing sufficient entropy are considered in both preceding and following timeslots, making it possible to retain them. 
\\
\textbf{Quantization.}
\label{par:sysdes-qs}
\textit{Quantization} translates a sensor signal (e.g., acceleration) to fingerprint bits used in a key agreement protocol. 
To ensure security, the produced fingerprints must be sufficiently unpredictable. 
Prior work~\cite{Bruesch:2019, Fomichev:2019perils} finds that quantization methods of state-of-the-art \gls{zip} schemes generate fingerprints with patterns (e.g., containing more 0-bits). 
We design the quantization of \name with three goals in mind: it must (1) generate fingerprints that are random, (2) apply across various sensors, and (3) reveal minimum information about the input sensor signal.
The last goal seeks to reduce adversary's knowledge about the input sensor signal leaked by quantization (e.g., signal range~\cite{Groza:2012}).

We quantize short sensor signals of several seconds, producing fingerprints of a few dozen bits. However, our method generalizes to longer signals and fingerprints. 
The advantage of using short input signals is twofold: (1) it forces an adversary to guess context captured by a sensor within a precision of a few seconds, (2) it requires less processing, improving the runtime performance of \name. 
Our quantization takes a sensor signal $S$ of length $N$ samples as input and outputs a fingerprint $f$ of $M$ bits (cf.~\autoref{sf:qs}). 
Specifically, we first find a quantization threshold $Thr_{Q}$ (solid red line in~\autoref{sf:qs}) that splits the signal horizontally into upper and lower parts. 
The threshold is computed from the median of the signal, ensuring that the same number of samples lie above and below it. 
This way of selecting $Thr_{Q}$ leads to improved randomness of the fingerprints (cf.~\autoref{sub:eval-fp-rand}) and is efficient to implement.
Second, we place the quantization points $p_{1},\dotsc, p_{M}$ (marked as \ding{58} in~\autoref{sf:qs}) equidistantly onto the threshold line within distance $\Delta_{Q} =  \ceil*{\frac{N}{M}} + \varepsilon$ from each other, covering the signal completely.
The number of quantization points and $\Delta_{Q}$ are public parameters customized for each sensor modality (cf.~\autoref{sub:eval-method}).
Using public parameters has the advantage of (1) fewer communication rounds, as they do not need to be exchanged during the pairing protocol, and (2) not leaking information about a specific input signal, as the parameters are derived from the class of signals of the same modality, and thus are general.  
Third, we obtain a bit in the fingerprint $f(i)$ by comparing a signal value at the quantization point $S(p_{i})$ with the quantization threshold $Thr_{Q}$:
\[
f(i) =
  \begin{cases}
    1, &  S(p_{i}) > Thr_{Q} \\
    0, &  \text{otherwise}
  \end{cases}
\]
\\
\textbf{Key Exchange.}
The main challenge of adapting \gls{fpake} for the use in \name is to find which minimum fingerprint sizes\footnote{Here, we assume fingerprints to be uniformly random bitstrings; entropy biases will increase fingerprint sizes (cf.~\autoref{sub:eval-fp-rand}).} are required to securely exchange a 128-bit key, protecting against offline attacks. 
With \gls{fpake} we can choose \textit{arbitrarily small} fingerprints, which are then amplified to match the size of the encoded secret (cf.~\autoref{fig:fpakeschem-1}). 
By reducing the number of required fingerprint bits, we are able to shorten pairing time, while providing sufficient security. 

Before calculating the required fingerprint sizes, we determine sufficient security levels for \name. 
Specifically, we consider two levels: (1) the minimal probability $P$ with which an offline attack is eliminated and (2)  the average complexity $C$ of an offline attack.  
We set $P=1-2^{-20}$, namely an adversary \textit{actively participating} in a million pairing sessions can mount an offline attack in at most one of them (without even learning which one). 
This level of security is considered adequate for \gls{zip}~\cite{Miettinen:2018}. 
We set $C=2^{60}$, demanding that an offline attack has an average complexity of at least $2^{60}$ AES decryptions.
We consider this complexity sufficient given that attack time is limited to few seconds due to the key confirmation timeouts that we augment the \gls{fpake} protocol with (cf.~\autoref{fig:fpakeschem-1}).

We explain how to calculate the required fingerprint sizes, satisfying our chosen security levels using a 95\% similarity threshold as an example. 
The similarity threshold defines the amount of common bits in two fingerprints needed to obtain a shared secret key. 
One might think to set the fingerprint size $|f|$ such that the probability of guessing at least $0.95\cdot|f|$ bits correctly is smaller than $2^{-20}$, since the key should be undecodable otherwise.
Unfortunately, this is not true: an \gls{ecc} (used in \gls{fpake}) correcting $5\%$ mismatch between the fingerprints \textit{leaks some information about the encoded secret} until up to $2\cdot 5\%=10\%$ mismatch.
Thus, an active adversary guessing less than 90\% of the fingerprint correctly learns nothing about the secret. However, if the guessed fingerprint is ``close enough''  (i.e., 90--95\% of the bits), the adversary cannot immediately decode the secret but obtains an \textit{ambiguous encoding} from which the secret can be brute-forced.
Taking this ``security gap'' inherent to \glspl{ecc} into account, we set $|f|$ such that the probability $\sum_{i=m}^{n}{n \choose i}/2^{n}$ of guessing $m=(2 \cdot Thr-1)\cdot|f|$ out of $n=|f|$ bits correctly is smaller than $2^{-20}$, where $Thr$ is the target similarity threshold.

\begin{table}
\small
	\begin{center}
 	\caption{Offline protection and brute-force complexity for \name computing $128$-bit keys for different choices of similarity thresholds and fingerprint sizes. Gray boxes mark sufficient security levels. $T$ is described in text.}
	\label{tab:fingerprintsizes}
	\begin{tabular}{cccc}
		\toprule
		\multirow{2}{*}{\makecell{Similarity \\ Threshold}} & 
		\multirow{2}{*}{\makecell{Fingerprint\\ Bits}} & 
		\multirow{2}{*}{\makecell{Offline Attack \\ $1-P$}} & 
		\multirow{2}{*}{\makecell{Brute-force \\ Complexity $C$}} \\ \\
		\midrule
		$95\%$ & 40 & \colorbox{lightgray!50!white}{$<2^{-23}$} & $\approx 2^{37}T$ \\ 
		$90\%$ & 60 & \colorbox{lightgray!50!white}{$<2^{-20}$} & $\approx 2^{32}T$\\   
		$85\%$ & 80 & $<2^{-12}$ & \colorbox{lightgray!50!white}{$\approx 2^{60}T$}\\   
        $80\%$ & 100 & $<2^{-7}$ & \colorbox{lightgray!50!white}{$\approx 2^{63}T$}\\   
        $75\%$ & 120 & $\approx 1$ & \colorbox{lightgray!50!white}{$\approx 2^{64}T$}\\ 
        $70\%$ & 140 & $\approx 1$ & \colorbox{lightgray!50!white}{$\approx 2^{60}T$} \\
		\bottomrule
	\end{tabular}
	\end{center}
	\end{table}

We note that $|f|$ goes to infinity when $Thr$ approaches $75\%$, since $2\cdot 25\%=50\%$ of a random bitstring is easy to guess.
Thus, full protection against offline attacks with probability at least $1-2^{-20}$ is only possible for thresholds over $75\%$, requiring short fingerprints of 40--60 bits for thresholds above $90\%$ (cf.~\autoref{tab:fingerprintsizes}).
Below $90\%$, the security of \name relies on our other security level measuring brute-force complexity of the offline attack.
For estimating this complexity, we think of \gls{ecc} encodings as consisting of $n=|f|$ parts which are correct or wrong depending on whether the corresponding fingerprint bit was correct or not.
We use the following brute-force method to decode an ambiguous encoding: randomly guess which $m$ parts of the encoding are correct, decode only them, and set the secret to be the first result that appears twice.
Considering that we do not know how many parts $i$ of the codeword are actually correct, 
the conditional probability of guessing $m$ out of $i$ correct parts in the $n$ parts long encoding is given by the hypergeometric distribution as $i^{\underline{m}}/n^{\underline{m}}$, which finds its maximum at $i=(n-m)/2$.
The complexity of the offline attack is thus lower bounded by $n^{\underline{m}}/(n-m)^{\underline{m}}T$. Here, $T$ is the complexity of $\mathsf{ECC.Decode}$, which is larger than the complexity of one AES decryption.

\autoref{tab:fingerprintsizes} shows the calculated fingerprint sizes providing sufficient security for \name based on \gls{fpake} for a range of similarity thresholds. 
In~\autoref{sec:discuss}, we elaborate that these findings are generic, thus can be directly reused by other \gls{zip} schemes. 
\section{Intra-car Device Pairing}
\label{sec:case-study}
We present \textit{intra-car device pairing}---an exemplary use case of \name to pair devices inside a moving car.
It enables novel vehicular applications such as pairing user devices for customized driving experience or pairing  \glspl{ecu} for travel efficiency~\cite{Claburn:2018, Sanz:2016}. We first provide the case overview followed by implementation details.
\\
\textbf{Case Overview.}
There is a growing number of on-board smart devices in modern cars, including devices of drivers and passengers (e.g., smartphone, earbuds) as well as \glspl{ecu} and infotainment systems~\cite{Claburn:2018, Iron:2018}. The prohibitive user effort to pair these devices, many of which lack user interfaces, justifies the use of \name for intra-car device pairing.
\name utilizes four sensor modalities to capture the context of a moving car: vertical and horizontal acceleration, gyroscope sky-axis, and barometer.
Our review of prior work shows that acceleration of a moving car can be decomposed into \textit{vertical} and \textit{horizontal components}, with the former capturing road conditions (i.e., bumpiness), while the latter---driving patterns and traffic conditions (i.e., acceleration/deceleration)~\cite{Chen:2017}. A \textit{gyroscope} measures car's turns and steering directions~\cite{Vaas:2018}, while a \textit{barometer} captures altitude changes when a car moves along the road~\cite{Sankaran:2014}. 
 
\subsection{Implementation}
\label{sec:impl}
\textbf{Data Collection.}
We develop an Android app to collect accelerometer, gyroscope, and barometer data at fixed sampling rates (i.e., 100 Hz for accelerometer and gyroscope; 10 Hz for barometer). 
We convert accelerometer and gyroscope data to the world coordinates, eliminating the effect of device orientation.
Before data collection, we perform an \gls{ntp} update on smartphones, ensuring consistent data timestamps, which we use to synchronize the start of sensor recordings of colocated devices.  
\\
\textbf{Data Processing.}
\label{subsub:data-proc}
We process the collected sensor data before feeding it into the activity filter. 
Prior to any processing, we resample the data to the set sampling rates, eliminating the effect of \textit{sampling rate instability}~\cite{Stisen:2015}. 
To decompose acceleration into vertical and horizontal components, we (1) remove the Earth's gravity from the accelerometer data applying a non-overlapping 5-second sliding window and (2) use the estimated Earth's gravity to perform the decomposition~\cite{Chen:2017}. 
For the gyroscope data, transformed to the world coordinates by our app, we select a Z-axis that is perpendicular to the road surface. We convert the barometer data $p$ to altitude $h_{alt}$ in meters using a standard pressure-height formula~\cite{Sankaran:2014}: 
\[ h_{alt} =  44330\cdot\biggl(1 - \left(\dfrac{p}{1013.25}\right)^{\tfrac{1}{5.255}}\biggr) \]

After converting the sensor data to a required format, we perform signal smoothing and noise reduction in two steps: (1) applying them on the whole data and (2) on signals of several seconds, partitioning these data. 
To remove low-frequency noise and smooth the whole data without distorting it (e.g., keep peak locations), we use a \gls{sg} filter with a window length 3 and degree 2 polynomial. Afterwards, we apply a Gaussian filter with a sigma of 1.4 to reduce high-frequency noise. 

We use the same sequence of filters on sensor signals of several seconds: the \gls{sg} filter has a window length 5 and degree 3 polynomial for finger-grained smoothing, while the Gaussian filter stays the same. 
For the acceleration signals, we afterwards apply an \gls{ewma} filter to smooth them further, while keeping their significant changes; the \gls{ewma} alpha is set to 0.16 and 0.2 for vertical and horizontal acceleration, respectively.
For the altitude signals, we perform mean subtraction before filtering. 
This helps to (1) remove offset between barometer sensors caused by hardware and temperature variation~\cite{Fomichev:2019}, (2) eliminate atmospheric pressure, accentuating altitude changes in the signal. 
We adapt filter parameters for signal smoothing and noise reduction from related work~\cite{Chen:2017, Han:2018, Lin:2019}.
\\
\textbf{Activity Filter.}
The activity filter applies to a processed sensor signal of several seconds. We implement it by computing the average power and \gls{snr} for all modalities, and counting prominent peaks for vertical and horizontal acceleration (cf.~\autoref{sec:sysdesign}). To pass the activity filter, a signal must have the average power, \gls{snr}, and optionally the number of prominent peaks higher than a predefined threshold, which we find empirically. The signal that passes the activity filter is input to the quantization, otherwise it is discarded. 
\\
\textbf{Quantization.}
We implement quantization, converting a sensor signal to fingerprint bits, as described in~\autoref{sec:sysdesign}. Its parameters (i.e., signal length, number of output bits) are set empirically for each modality (cf.~\autoref{sub:eval-method}). We compute the quantization threshold as a median of the sensor signal; for vertical and horizontal acceleration, we add small $\Delta$ to the median, reducing the effect of sensor noise on quantization. 
We concatenate  bits quantized from the sensor signal with bits derived from other modalities likewise before inputting them as one fingerprint to the \gls{fpake} protocol.  
\\
\textbf{\name Prototype.}
\label{subsub:prot}
We implement \name to evaluate its runtime performance on off-the-shelf \gls{iot} devices (cf.~\autoref{sub:eval-prot}).
We focus on the \gls{fpake} protocol, as the underlying functionality takes either constant (e.g., sensing) or negligible time (e.g., quantization).
The \name prototype allows two devices with similar fingerprints to establish a shared symmetric key. 
Our implementation is modular and agnostic to the fingerprint derivation, making it directly reusable by other \gls{zip} schemes. 
To implement the \gls{fpake} protocol, we use primitives from a Python cryptography library~\cite{Pyca:2020}.
For the \gls{ecc}, we utilize Shamir's secret sharing scheme in its error-correcting variant (i.e., introducing redundancy by adding more point-value pairs of the polynomial)~\cite{Fpake:2018}.
For the PAKE component, we use the \gls{eke} protocol~\cite{Bellovin:1992}, built as Diffie-Hellman key exchange symmetrically encrypted with passwords. 
Our \gls{fpake} implementation supports two security levels, generating keys of 128- and 244-bits. 
We enable communication between devices utilizing IP sockets and data serialization~\cite{Py-socket:2020, Py-pickle:2020}, allowing us to run the \name prototype in real-time.   
To benchmark our implementation, we employ a Python time module~\cite{Py-time:2020}. 

\section{Evaluation}
\label{sec:eval}
\begin{figure}
	\begin{center}
		\includegraphics[width=0.96\linewidth]{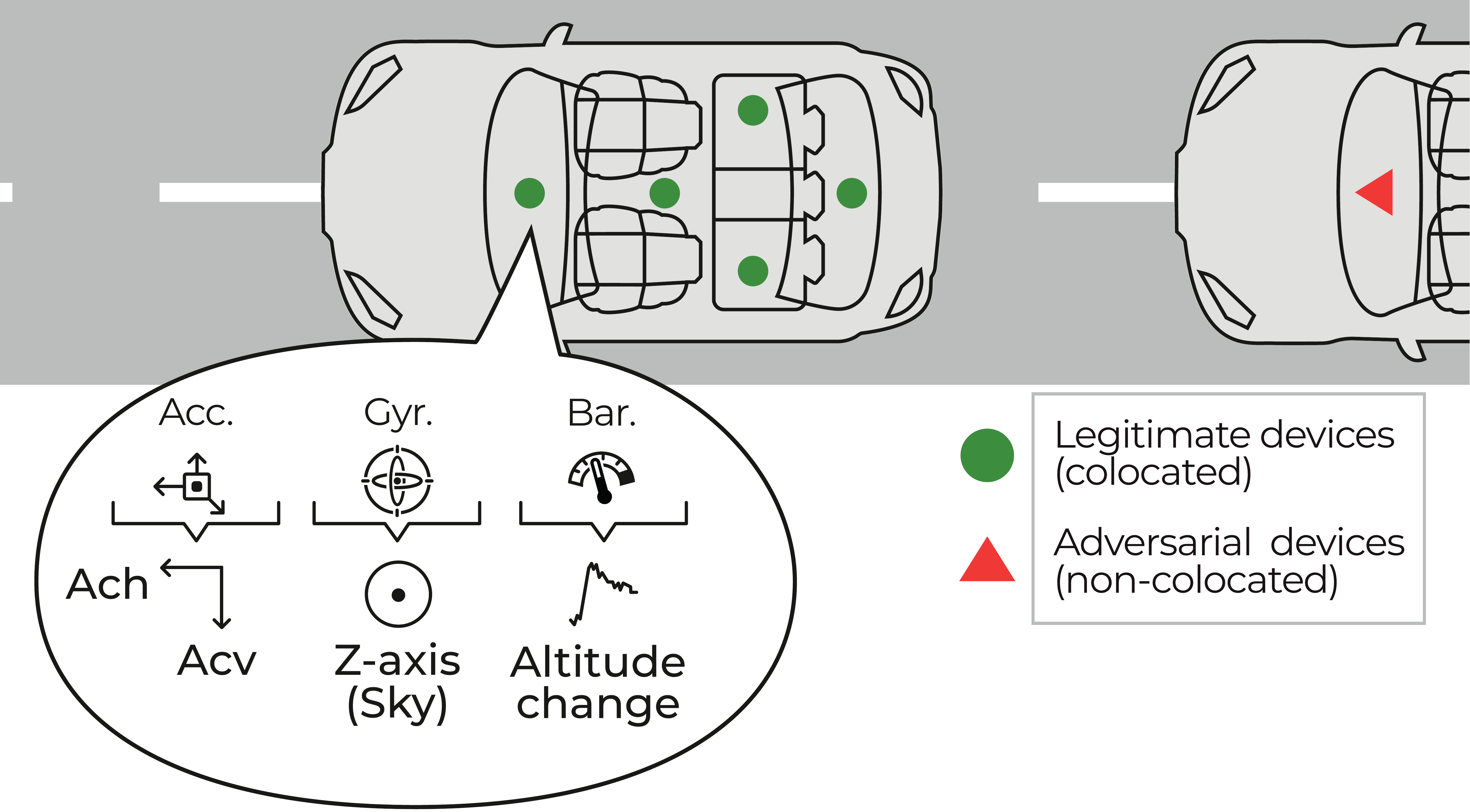}
	\end{center}
	\caption{Experiment setup. Smartphones are placed at five various spots inside each of two cars driving the same route.}
	\label{fig:car}
\end{figure}

We present a comprehensive evaluation of \name based on the real-world data we collect. 
\\
\textbf{Experiment Setup.}
\label{sub:eval-setup}
We collect accelerometer, gyroscope, and barometer data from four cars driven in a number of scenarios: within a \textit{city}, on \textit{country} roads, on a \textit{highway}, and inside a \textit{parking} garage for a total of 800 km. 
To evaluate the suitability of \name for various cars, we collect these data using (1) two \textit{similar cars} (Opel Astra wagons; 400 km of driving) and (2) two \textit{different cars} (\v{S}koda Octavia sedan and Volkswagen Golf hatchback; 400 km of driving). 
In both experiments, we equip two cars with five smartphones each, covering spots where smart devices are typically found~\cite{Chen:2017, Fomichev:2019perils}: on a dashboard, between front seats, behind driver and passenger seats, and inside a trunk (cf.~\autoref{fig:car}). 
Then, we collect sensor data from two cars driven as such: (1) one car starts a predefined route, followed by another car after a 10--15 minute lag, (2) two cars drive one after another, changing the distance between each other and the role of a leading vehicle. We cover a similar number of kilometers for \textit{city}, \textit{country}, and \textit{highway} driving. In the \textit{parking} scenario, cars leave an underground garage and return back to it multiple times. To collect the sensor data, we utilize Nexus 5X and Nexus 6P smartphones.
After data processing (cf.~\autoref{subsub:data-proc}), we use vertical and horizontal acceleration (labeled as \textit{Acv} and \textit{Ach}, respectively), gyroscope sky-axis (\textit{Gyr}), and altitude computed from barometer (\textit{Bar}) in our evaluation. 
\\
\textbf{Reproducibility and Reusability.}
We release the collected sensor dataset along with the driven routes map and the source code of our data collection app, evaluation stack, and \name prototype~\cite{Subsample:2021}. 

\subsection{Methodology}
\label{sub:eval-method}
We evaluate \name using several criteria: (1) security and usability, (2) pairing time, and (3) runtime performance.
To assess security, we compute \gls{far} and evaluate entropy of our fingerprints. 
A false acceptance occurs when non-colocated devices in different cars (cf.~\autoref{fig:car}) pair because their fingerprints are similar enough. 
We assess usability by computing \gls{tar}, showing the rate of successful pairings between colocated devices inside the same car. 
For a detailed analysis of \glspl{far} and \glspl{tar}, we compute them on the \textit{full} data of an experiment (e.g., \textit{similar cars}) and on the subsets of data corresponding to driving in one of our scenarios: \textit{city}, \textit{country}, \textit{highway}, and \textit{parking}. 
To evaluate pairing time, we find the amount of context data (in seconds) required to pair securely, while for runtime performance we benchmark the \name prototype on the Raspberry Pi.
\\
\textbf{System Parameters.}
\label{para:sys-params}
We use the collected sensor data to find configuration parameters for \name's modules: activity filter, quantization, and \gls{fpake} yielding the best trade-off between security and short pairing time.
To find the \textit{length of sensor signal} to derive fingerprint bits, we examine how much sensor data is required to capture typical ambient activity (e.g., car turn by \textit{Gyr}).
Our results show that 10 seconds of \textit{Acv}, \textit{Ach}, and \textit{Gyr} data capture typical road bumpiness, acceleration patterns, and car turns, while 20 seconds of \textit{Bar} data is enough to record  altitude changes. 
We set these signal lengths as input to the activity filter and to quantization, using them to empirically find thresholds for activity filter metrics for each sensor modality.

To choose the number of \textit{fingerprint bits} output by quantization, we investigate (1) the good ratio between high \gls{tar} and low \gls{far} and (2) modality variation. The latter helps us understand how many uncorrelated bits can be extracted from the sensor signal. Based on our findings, we set the number of fingerprint bits to 24 for both \textit{Acv} and \textit{Ach}, 16 for \textit{Gyr}, and 12 for \textit{Bar}.
A \textit{similarity threshold} defines the level of similarity between two fingerprints required to establish pairing. 
To select similarity thresholds, we study how many bits typically differ in the fingerprints of colocated devices. 
We set the following thresholds, balancing high \gls{tar} and low \gls{far}, to be used in the \gls{fpake} protocol:
70.8\% (\textit{Acv}), 75\% (\textit{Ach}), 93.7\% (\textit{Gyr}), and 91.7\% (\textit{Bar}). 

\begin{figure*}
	\centering
	\begin{subfigure}[b]{.26\textwidth}
		\centering
		\includegraphics[width=\linewidth]{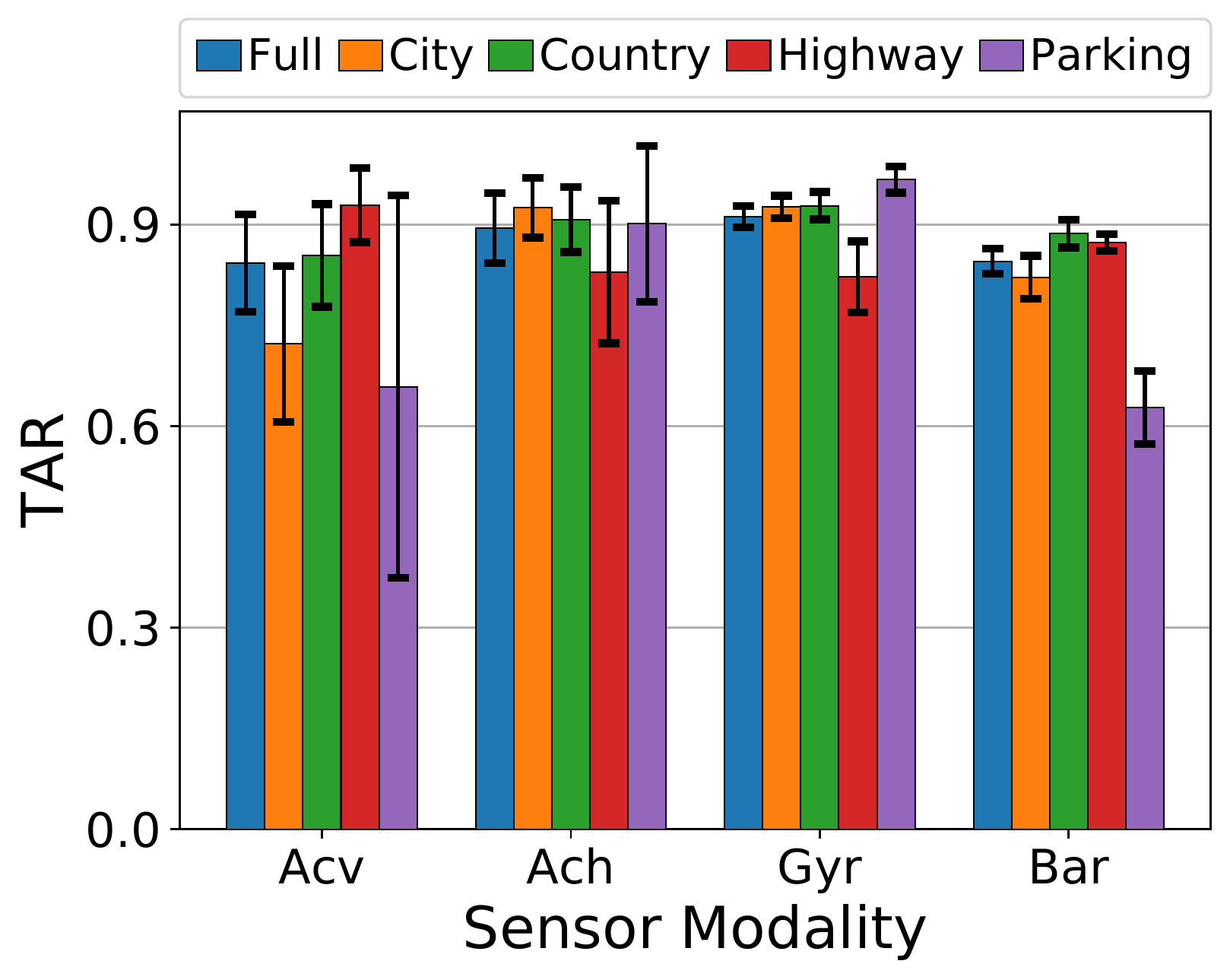}
		\caption{\glspl{tar}}
		\label{sf:tar-indiv}
  	\end{subfigure}
  	\begin{subfigure}[b]{.26\textwidth}
		\centering
		\includegraphics[width=\linewidth]{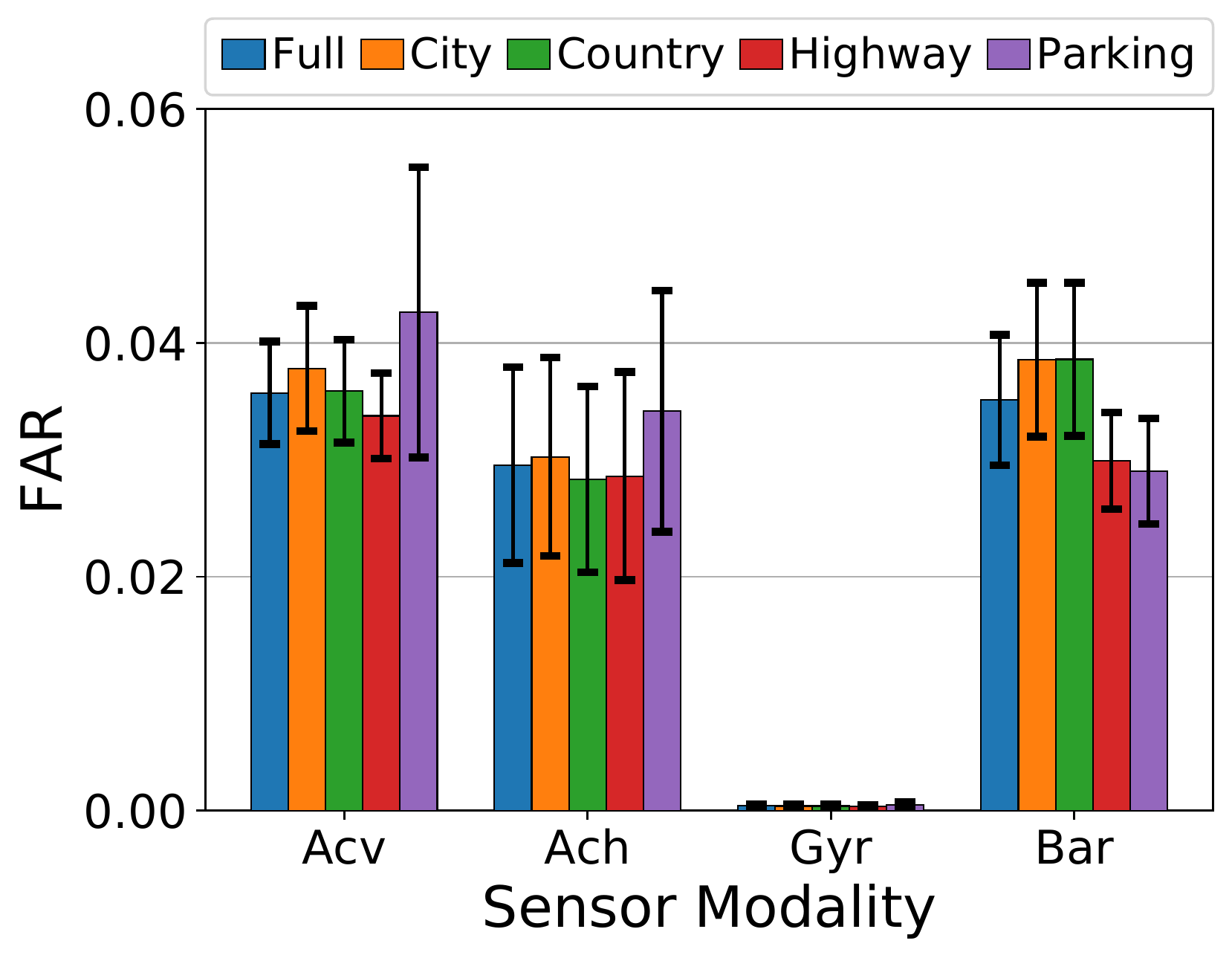} 
		\caption{\glspl{far}: injection attack}
		\label{sf:far-indiv-inject}
  	\end{subfigure}
	\begin{subfigure}[b]{.26\textwidth}
		\centering
	    \includegraphics[width=\linewidth]{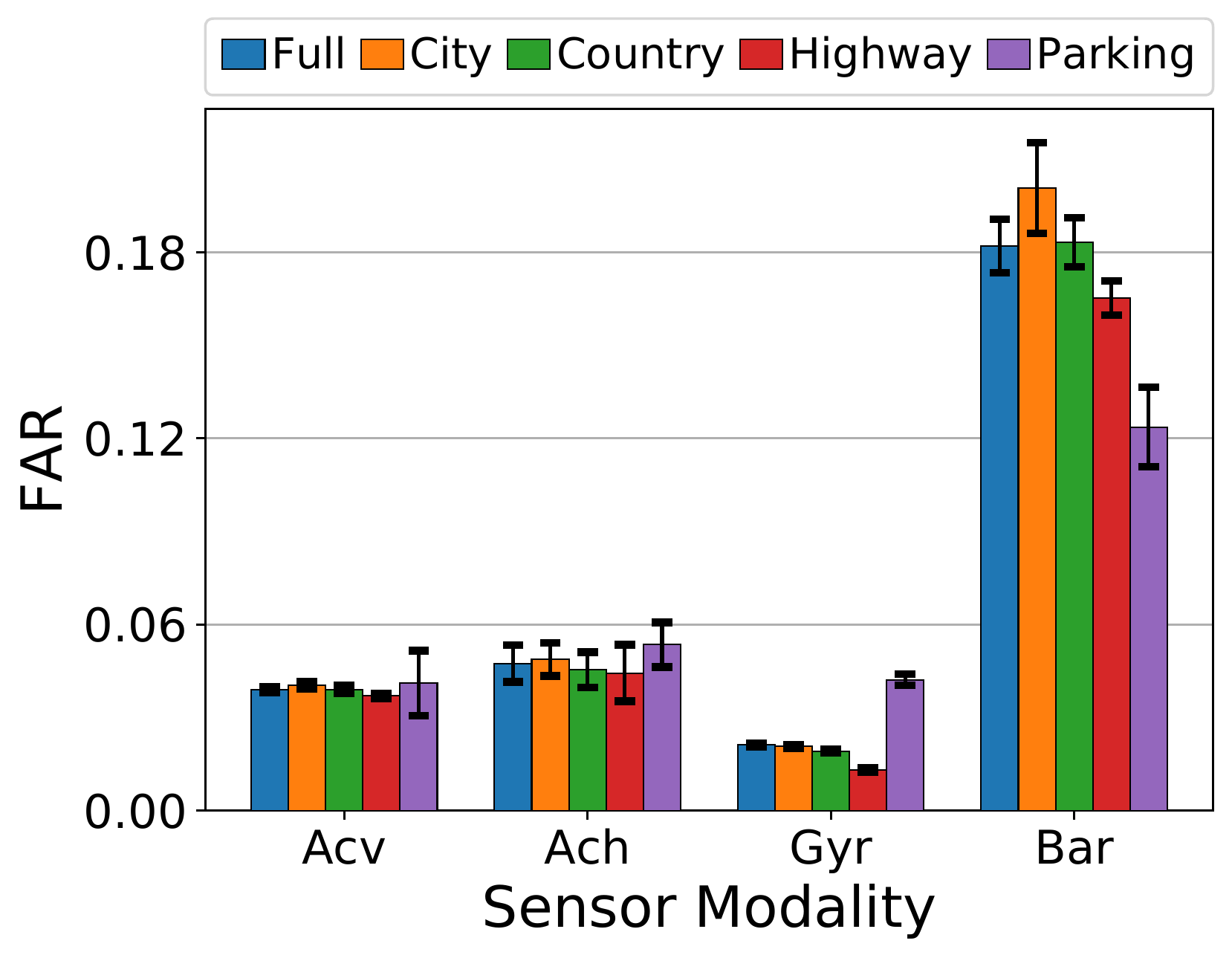}
	    \caption{\glspl{far}: replay attack}
	    \label{sf:far-indiv-replay}
	\end{subfigure}
	\caption{True Acceptance Rates (TARs) and False Acceptance Rates (FARs) for individual sensors.}
	\label{fig:indiv-error-rates}
\end{figure*}

\begin{figure*}
	\centering
	\begin{subfigure}[b]{.245\textwidth}
		\centering
		\includegraphics[width=\linewidth]{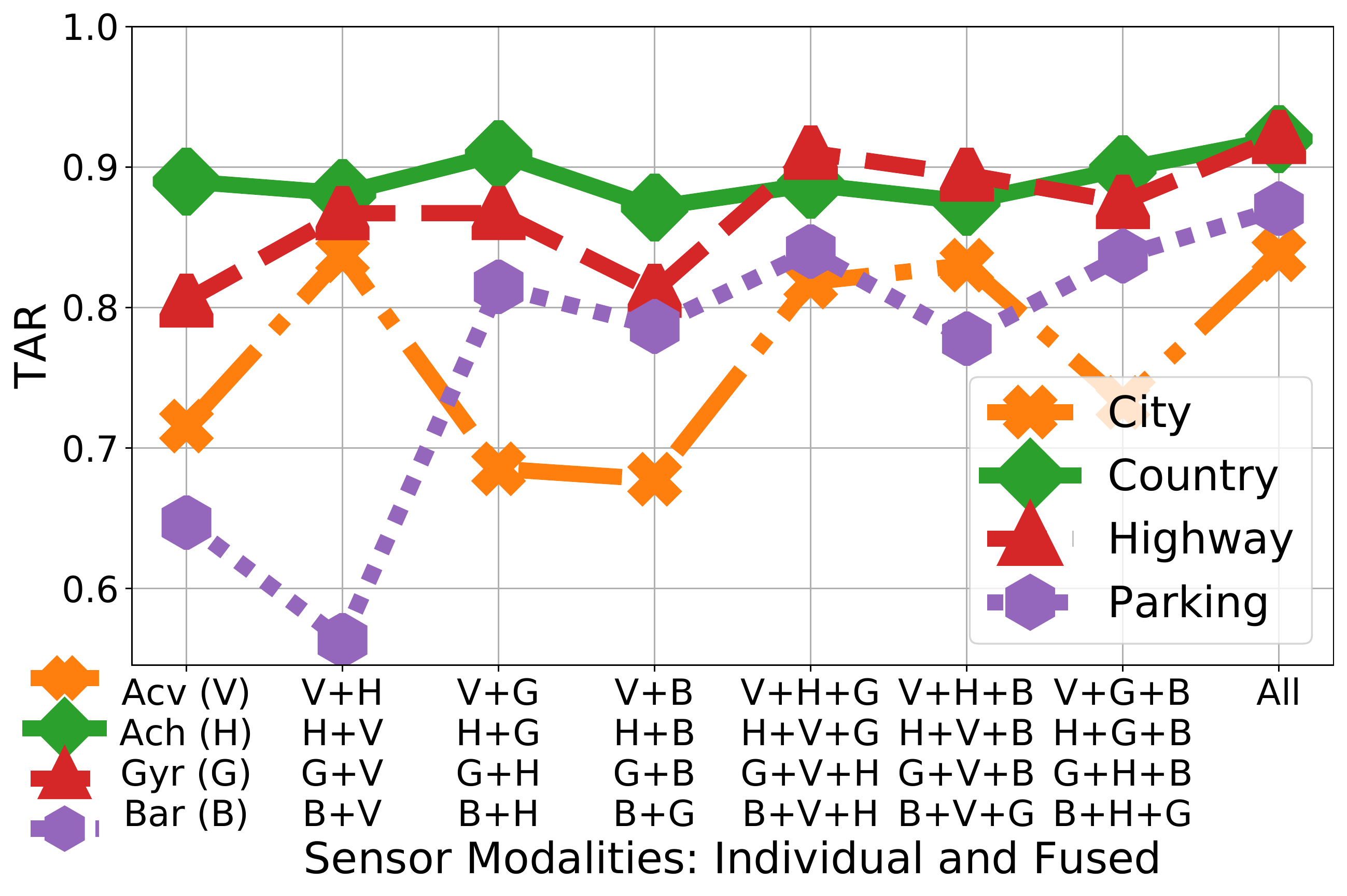}
		\caption{\glspl{tar}}
		\label{sf:tar-fused}
  	\end{subfigure}
  	\begin{subfigure}[b]{.245\textwidth}
		\centering
		\includegraphics[width=\linewidth]{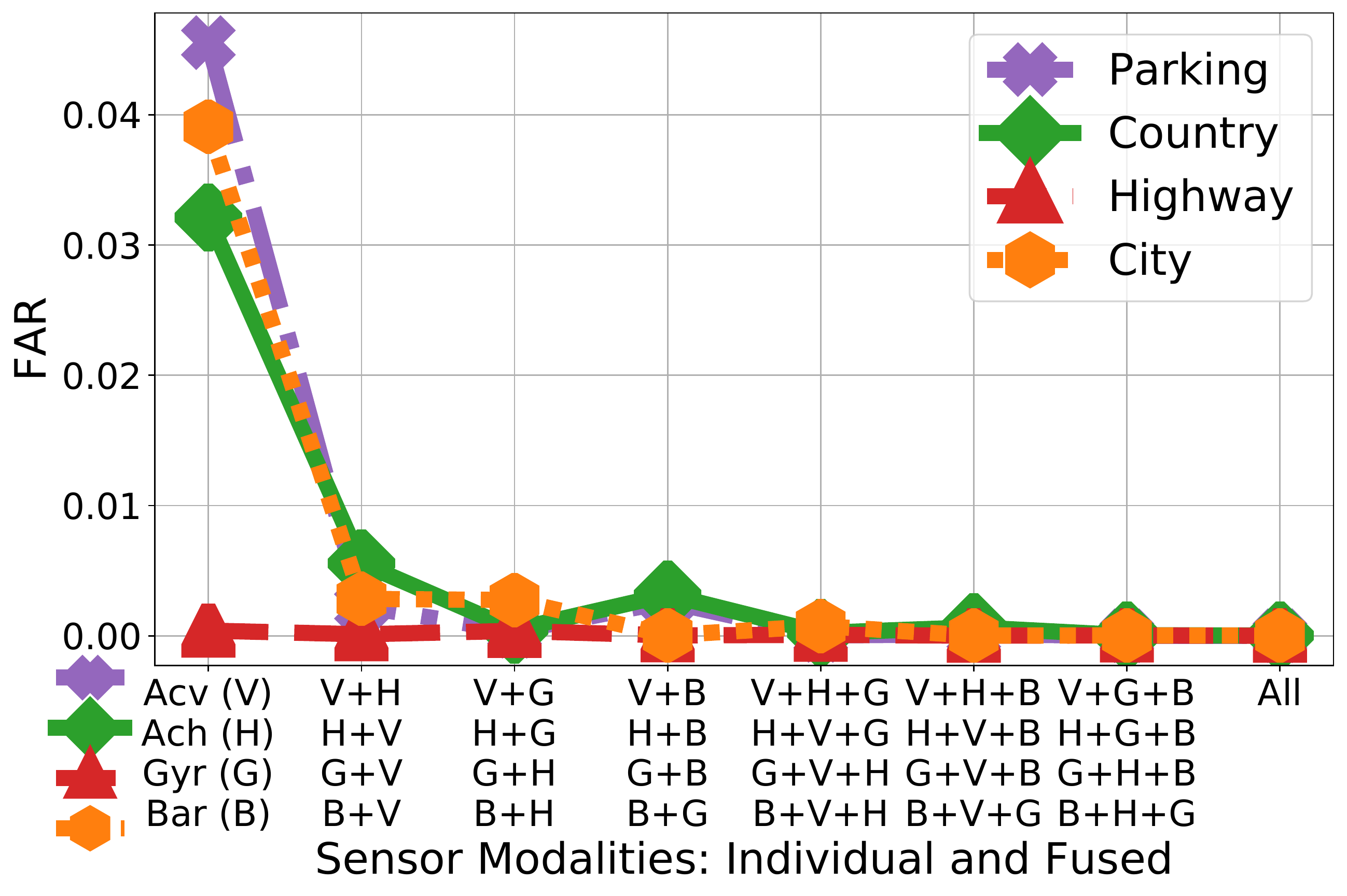} 
		\caption{\glspl{far}: injection attack}
		\label{sf:far-fused-inject}
  	\end{subfigure}
	\begin{subfigure}[b]{.245\textwidth}
		\centering
	    \includegraphics[width=\linewidth]{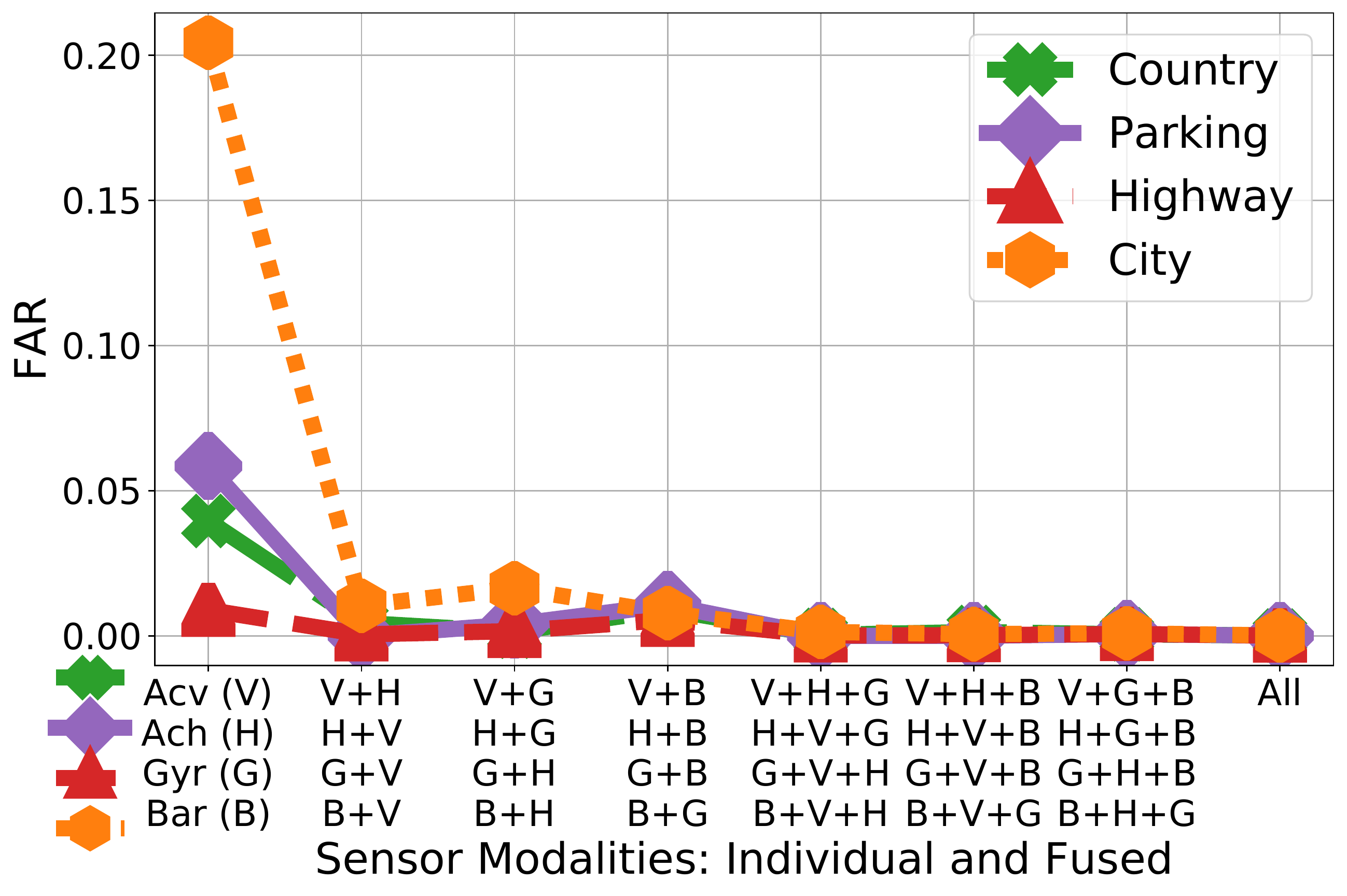}
	    \caption{\glspl{far}: replay attack}
	    \label{sf:far-fused-replay}
	\end{subfigure}
    \begin{subfigure}[b]{.245\textwidth}
		\centering
	    \includegraphics[width=\linewidth]{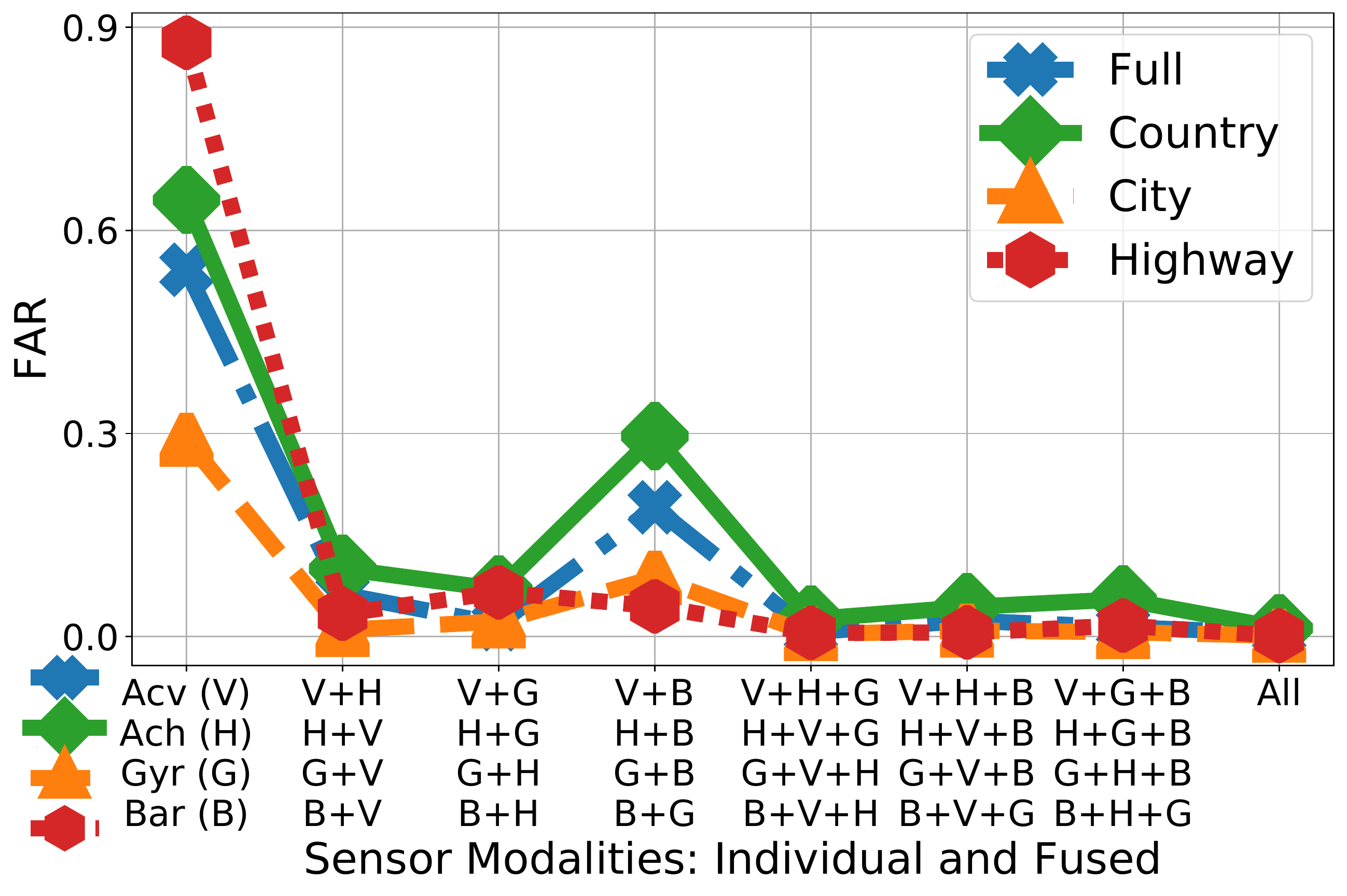}
	    \caption{\glspl{far}: similar-context attack}
	    \label{sf:far-sim-context}
	\end{subfigure}
	\caption{Effect of sensor fusion on True Acceptance Rates (TARs) and False Acceptance Rates (FARs) for representative subsets of scenarios; we connect markers for readability, the plots do not show a time series.}
	\label{fig:fused-error-rates}
\end{figure*}

\subsection{Pairing between Colocated Devices}
\label{sub:eval-benign}
We compute \glspl{tar} between each pair of colocated devices inside the same car, providing the average \gls{tar}. First, we present \glspl{tar} for individual sensors (e.g., \textit{Acv}) followed by the evaluation of sensor fusion. Our results are consistent across the \textit{similar} and \textit{different cars} experiments, indicating generalizability of \name to various cars. In the following, we provide typical \glspl{tar}.   
\\
\textbf{\glspl{tar} of Individual Sensors.}
\autoref{sf:tar-indiv} depicts \glspl{tar} for the first car in the \textit{similar cars} experiment. 
We see that \textit{full} \glspl{tar} range between 0.84 and 0.91, showing that the individual sensors alone achieve relatively high success rates. 
However, the \glspl{tar} of scenarios (e.g., \textit{city}) have higher variation: while \textit{Ach} and \textit{Gyr} exhibit fairly consistent \glspl{tar}, \textit{Bar} and especially \textit{Acv} show a wider spread of \glspl{tar}. 
For \textit{Acv}, the \gls{tar} spread is caused by diverse bumpiness perception inside a car affected by such factors  as car suspension (e.g., front vs. rear) and surface on which bumpiness is measured (e.g., plastic vs. fabric). 
These factors become important when a car moves slowly, reducing \glspl{tar} as in the \textit{city} and \textit{parking}, while higher speed leads to more profound bumpiness, increasing \glspl{tar} as in the \textit{country} and \textit{highway}.
For \textit{Bar}, higher speed causes profound altitude changes, improving \glspl{tar} (e.g., \textit{highway}),
while there are few such changes when traveling short distances, reducing \glspl{tar} (e.g., \textit{parking}). 
In contrast, \textit{Ach} and \textit{Gyr} show lower \glspl{tar} when a car moves at constant speed (e.g., \textit{highway}). 
These sensors benefit from non-monotonic driving with many stops, leading to distinct acceleration patterns (\textit{Ach}) and sharp turns (\textit{Gyr}), as in the \textit{city} and \textit{parking}.  
Thus, no sensor outperforms the others in all the scenarios, and they show potential for complementing each other.

We analyze \gls{tar} deviation inside a car, finding that longer distance between devices leads to lower \glspl{tar}.
This happens because context signals (e.g., road bumpiness) can be attenuated or perceived with varying intensity at distant spots. 
We find that rapidly changing sensors (i.e., \textit{Acv}, \textit{Ach}) can have up to 20 percentage points of \gls{tar} difference between farthest devices, while for gradually changing sensors (i.e., \textit{Gyr}, \textit{Bar}) it is below five percentage points. 
\\
\textbf{\glspl{tar} with Sensor Fusion.}
We fuse sensors by concatenating sub-fingerprints of different modalities derived in the same timeframe. 
Thus, we obtain more fingerprint bits in less time, speeding up pairing. 
We explore the fusion of two, three, and all of our sensors.
Our findings show that sensor fusion generally increases \glspl{tar}, while reducing their deviation between devices.
This happens because sensors can reinforce each other in the following way: error correction bits unused in the sub-fingerprint of highest similarity allow fixing extra errors in another sub-fingerprint, making the fused fingerprint exceed the similarity threshold, improving the \gls{tar}.
Such reinforcing effect leads to the fused \gls{tar} to be either close to the highest \gls{tar} in sensor combination or even exceed it.
The latter outcome is typical for sensor combinations including \textit{Ach} and \textit{Gyr}, which often capture co-occuring ambient activity (e.g., decelerate when turning).

\autoref{sf:tar-fused} shows a subset of fused \glspl{tar} for the second car in the \textit{similar cars} experiment. We see that by adding more sensors \glspl{tar} steadily increase from left to right: ranging from (0.65, 0.89) for individual modalities to (0.85, 0.93) when fusing all of them. 
With the \gls{tar} of 0.9 colocated devices would need 1.1 pairing attempts on average to pair successfully. 
In few cases, sensors do not reinforce each other, namely combinations including \textit{Acv} and \textit{Bar}, and \textit{Acv} and \textit{Gyr} in the \textit{parking} and \textit{city}, leading to reduced \glspl{tar}.
For \textit{Acv} and \textit{Bar}, both have lowest \glspl{tar} in these scenarios (cf.~\autoref{sf:tar-indiv}), so combining them increases the number of mismatching bits in the fused fingerprint.  
We find that \textit{Acv} and \textit{Gyr} often capture disjoint ambient activity (e.g., high speed: intense bumpiness but no turns), explaining lower potential for reinforcing each other.  

\subsection{Resilience to Attacks}
\label{sub:eval-adv} 
We compute \glspl{far} between each pair of non-colocated devices in different cars, presenting the average \gls{far} under \textit{injection}, \textit{replay}, and \textit{similar-context} attacks (cf.~\autoref{sec:mod}).
Similar to \gls{tar}, we first provide \glspl{far} for individual sensors and then evaluate their fusion, obtaining consistent results across \textit{similar} and \textit{different cars} experiments. In the following, we provide typical \glspl{far}. 
\\
\textbf{Injection Attack.}
\label{subsub:inject}
We use sensor data collected inside a parked car capturing noise to pair with legitimate devices. 
\autoref{sf:far-indiv-inject} depicts \glspl{far} of individual sensors computed for the first car in the \textit{similar cars} experiment.
We see that three out of four modalities show \glspl{far} above 0.03, making this low-effort attack practical. 
However, injecting sensor noise does not work on \textit{Gyr} because car turns result in distinct up and down peaks in the signal (cf.~\autoref{sf:qs}) that are not common for noise. 
With sensor fusion, \glspl{far} drop below half a percentage point using two modalities, converging to zero by adding more sensors (cf.~\autoref{sf:far-fused-inject}).
This result is the opposite of the reinforcing effect in \glspl{tar}, showing that with more sensors differences between non-colocated fingerprints grow, reducing \glspl{far}. 

We also try injecting sensor signals that are collected in a moving car but do not pass the activity filter. 
In this case, \glspl{far} grow by an extra percentage point for \textit{Acv}, \textit{Ach}, and \textit{Bar}, while for \textit{Gyr} they increase by order of magnitude: up to 0.005. 
Thus, low-entropy sensor signals from a moving car slightly improve the attack, while sensor fusion has the same effect as in~\autoref{sf:far-fused-inject}. 
\\
\textbf{Replay Attack.}
\label{subsub:replay}
We replay sensor signals passing the activity filter from one car to pair with devices in another car; both cars have driven the same route. 
In the first case, we do not synchronize such replayed signals. 
\autoref{sf:far-indiv-replay} depicts \glspl{far} of individual sensors for the first car in the \textit{similar cars} experiment. 
Compared to injection attack, \glspl{far} show a fourfold increase for \textit{Gyr} and \textit{Bar}, remaining similar for \textit{Acv} and \textit{Ach}.
The altitude change (\textit{Bar}) on a given route has least variation, allowing successfully replay (i.e., \gls{far} of up to 0.2), while other sensors are less affected. We can reach zero \glspl{far} by fusing more than two sensors (cf.~\autoref{sf:far-fused-replay}). 

In the second case, we replay sensor signals from periods when both cars drive the same part of the route (e.g., in a city) using a rough timeline of their travel. 
We see an extra twofold increase in \glspl{far} of \textit{Gyr} and \textit{Bar} peaking at 0.07 and 0.38, respectively, while for \textit{Acv} and \textit{Ach} the growth is 1--3 percentage points. 
Thus, all sensors have \gls{far} above 0.05, making this attack alarming. 
The sensor fusion leads to zero \glspl{far} as in~\autoref{sf:far-fused-replay}, showing its importance to prevent replay attacks. 
\\
\textbf{Similar-context Attack.}
\label{subsub:sim-context}
We use sensor signals passing the activity filter from one car to pair with devices in another car when two cars drive one after another (cf.~\autoref{fig:car}). 
We grant the adversary an unfair advantage of matching a single sensor (e.g., \textit{Acv}). It means that they always 
``guess'' the closest fingerprint to the legitimate one; the adversarial and legitimate fingerprints are derived from temporally close sensor signals. 
\autoref{sf:far-sim-context} depicts the best achievable \glspl{far} for this attack. 
We see that none of individual sensors can prevent the similar-context attack alone, showing \glspl{far} between 0.3 and 0.9 (leftmost of the graph). 
As in the replay, \textit{Bar} that has least variation is the most vulnerable followed by \textit{Ach} and \textit{Acv}. 
For \textit{Ach} and \textit{Acv}, \glspl{far} are caused by shared road conditions such speed limits leading to consistent decelerations (\textit{Ach}) and road cracks resulting in similar bumpiness (\textit{Acv}). 
\textit{Gyr} is the most robust to this attack because it captures human-specific steering behavior, which varies between drivers.  

We see that fusing two sensors cannot prevent the similar-context attack, especially when combining low-varying \textit{Bar} with other modalities (cf. peak in the middle of~\autoref{sf:far-sim-context}). 
By adding three and more sensors, we achieve nearly zero \glspl{far}, emphasizing the necessity for sensor fusion to mitigate advanced attacks in \gls{zip}. 

\subsection{Pairing Time}
\label{sub:eval-pair-time}
We compare pairing time of \name utilizing \gls{fpake} and state-of-the-art \gls{zip} schemes based on fuzzy commitments. 
To enable a fair comparison, we assume the number of fingerprint bits and time to derive them to be the same (cf.~\autoref{sub:eval-method}) and evaluate pairing time on the level of the cryptographic protocol, namely \gls{fpake} vs. fuzzy commitments. 
First, we calculate how much time it takes to obtain enough fingerprint bits to provide security against offline attacks for \gls{fpake} and fuzzy commitments. 
For the former, we use findings in~\autoref{tab:fingerprintsizes}, while for the latter we target a 128-bit fingerprint, accounting for entropy loss due to error correction~\cite{Miettinen:2018}: 
\[ |f|_{entropy\_loss} =  |f|_{target} + 2\cdot(1 - thr)\cdot|f|_{target} \]
Here, $thr$ denotes a similarity threshold. 
\autoref{tab:pairing-time} shows the resulting pairing times, demonstrating that \name requires 20--40 seconds to pair in the majority of cases, while state-of-the-art schemes need 1.5--3 times longer time under the same conditions.  

\begin{table}
\small
\centering
	\caption{Calculated pairing times for \name (\gls{fpake}) and state-of-the-art \gls{zip} schemes (Fuzzy commitments: F. com.).}
	\label{tab:pairing-time}
  \begin{tabular}{c|c|cc|cc}
  	\toprule
  	\multirow{2}{*}{\makecell{Sensor \\ (Fusion)}} & \multirow{2}{*}{\makecell{Sim. \\ Thr.}}  & \multicolumn{2}{c|}{Fingerprint Bits} & \multicolumn{2}{c}{Pairing Time (s)} \\
  	& & fPAKE & F. com. & fPAKE & F. com. \\
  	\midrule
  	Acv (V) & 70.8\% & 140 & 203 & 60 & 90\\ 
  	Ach (H) & 75.0\% & 120 & 192 & 50 & 80\\ 
  	Gyr (G) & 93.7\% & 50 & 145 & 40 & 100\\ 
  	Bar (B) & 91.7\% & 60 & 147 & 100 & 260\\ 
  	\hline
  	V+H & 72.9\% & 130 & 198 & 30 & 50\\ 
  	V+G & 80.0\% & 100 & 180 & 30 & 50\\ 
    V+B & 77.8\% & 110 & 185 & 80 & 120\\ 
    H+G & 82.5\% & 90 & 173 & 30 & 50\\ 
    H+B & 80.5\% & 100 & 178 & 60 & 100\\ 
    G+B & 92.9\% & 55 & 147 & 40 & 120\\ 
    \hline
    V+H+G & 78.1\% & 110 & 185 & 20 & 30\\ 
    V+H+B & 76.7\% & 120 & 188 & 40 & 80\\ 
    V+G+B & 82.7\% & 90 & 173 & 40 & 80\\ 
    H+G+B & 84.6\% & 80 & 168 & 40 & 80\\ 
    \hline
    All & 80.2\% & 100 & 179 & 40 & 60\\ 
  	\bottomrule
  \end{tabular}
\end{table}

\begin{figure}
\centering
  \includegraphics[width=\linewidth]{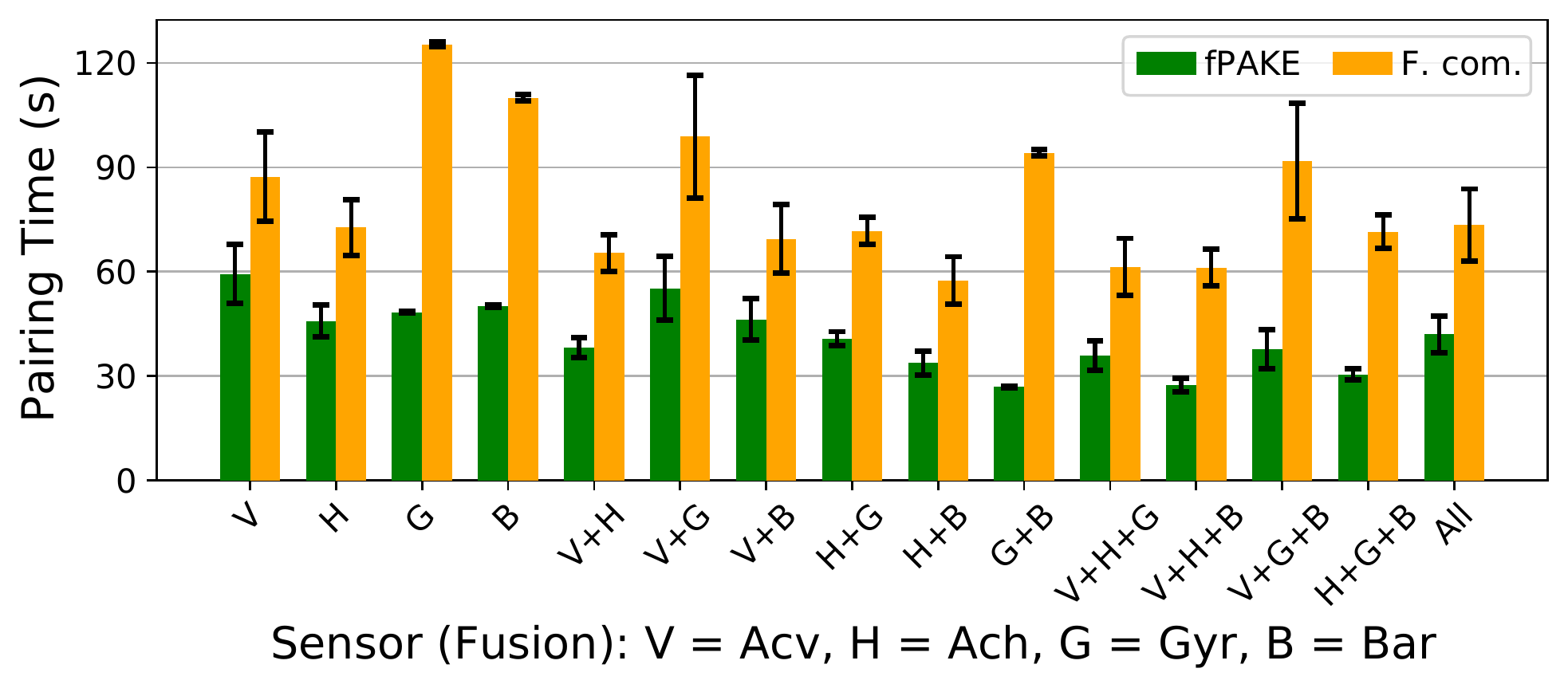}
\caption{Pairing times obtained from our sensor data for \name (\gls{fpake}) and state-of-the-art \gls{zip} schemes (Fuzzy commitments: F. com.).}
\label{fig:pair-time-data}
\end{figure}

Second, we evaluate the time required to accumulate fingerprint bits for \gls{fpake} and fuzzy commitments in~\autoref{tab:pairing-time} by traversing our collected sensor data with an overlapping sliding window using a 5-second step (cf. Activity Filter in~\autoref{sec:sysdesign} for reasoning).
~\autoref{fig:pair-time-data} gives pairing times obtained on the \textit{full} data of the first car in the \textit{different cars} experiment, confirming the 1.5--3 faster pairing time of \name. 
The calculated and obtained from our data pairing times are close to each other; the latter pairing times for \textit{Bar} and its fusion combinations are even smaller, as the length of the \textit{Bar} signal (i.e., 20 seconds) is significantly bigger than the sliding window step. 
We see that pairing times obtained from our data shorten in the case of profound ambient activity (e.g., \textit{Acv} on \textit{highway}), and pairing time consistency inside a car depends on device location for \textit{Acv} and \textit{Ach}, while it is stable for \textit{Gyr} and \textit{Bar}.

\begin{figure}
\centering
  \begin{minipage}[b]{.46\textwidth}
    \centering
		\subcaptionbox{Acv\label{sf:acv-rand}}{%
		\centering
			\includegraphics[width=0.42\textwidth]{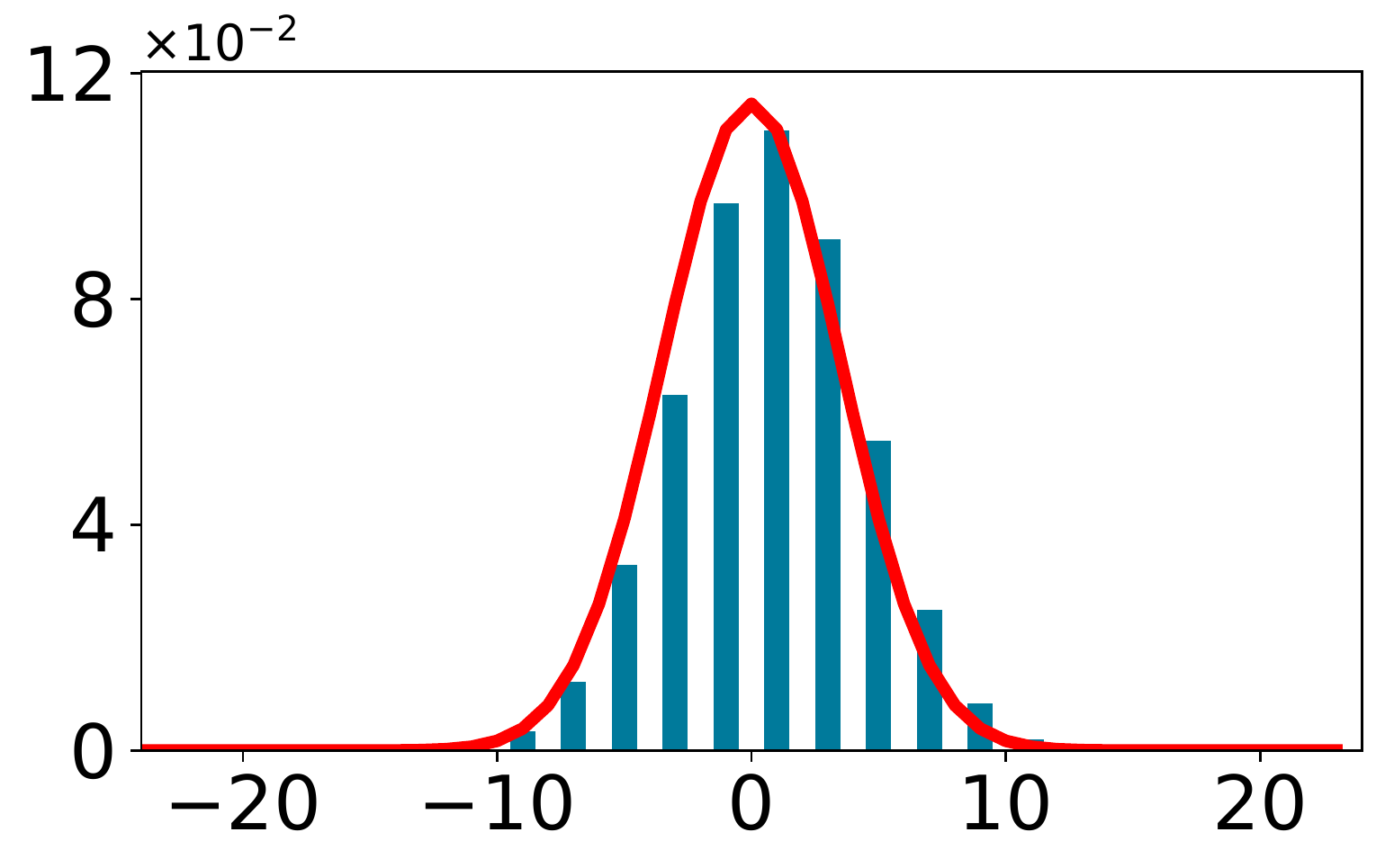}%
		}
		\subcaptionbox{Ach\label{sf:ach-rand}}{%
		\centering
			\includegraphics[width=0.42\textwidth]{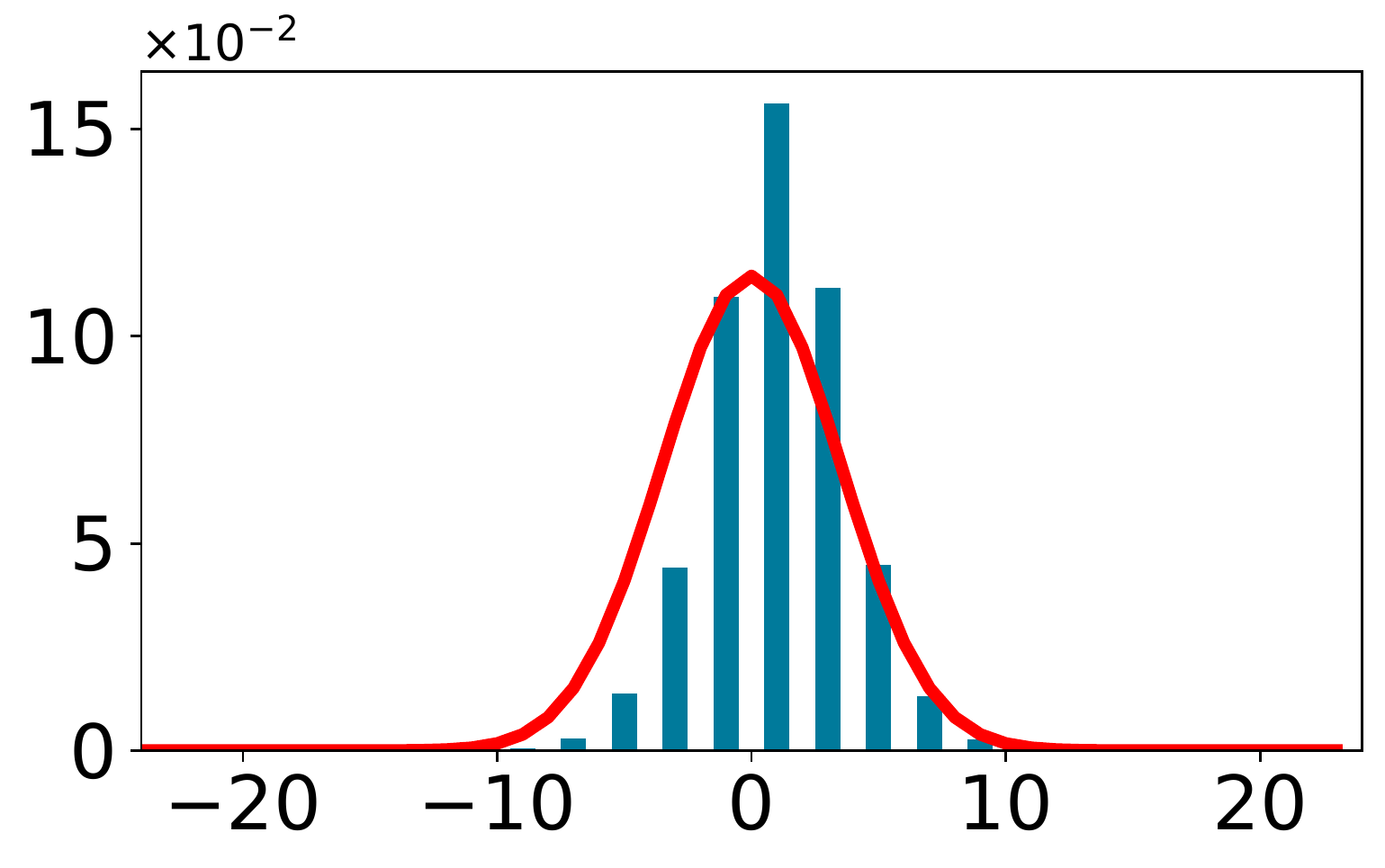}%
		}
	
		\subcaptionbox{Gyr\label{sf:gyr-rand}}{%
		\centering
			\includegraphics[width=0.42\textwidth]{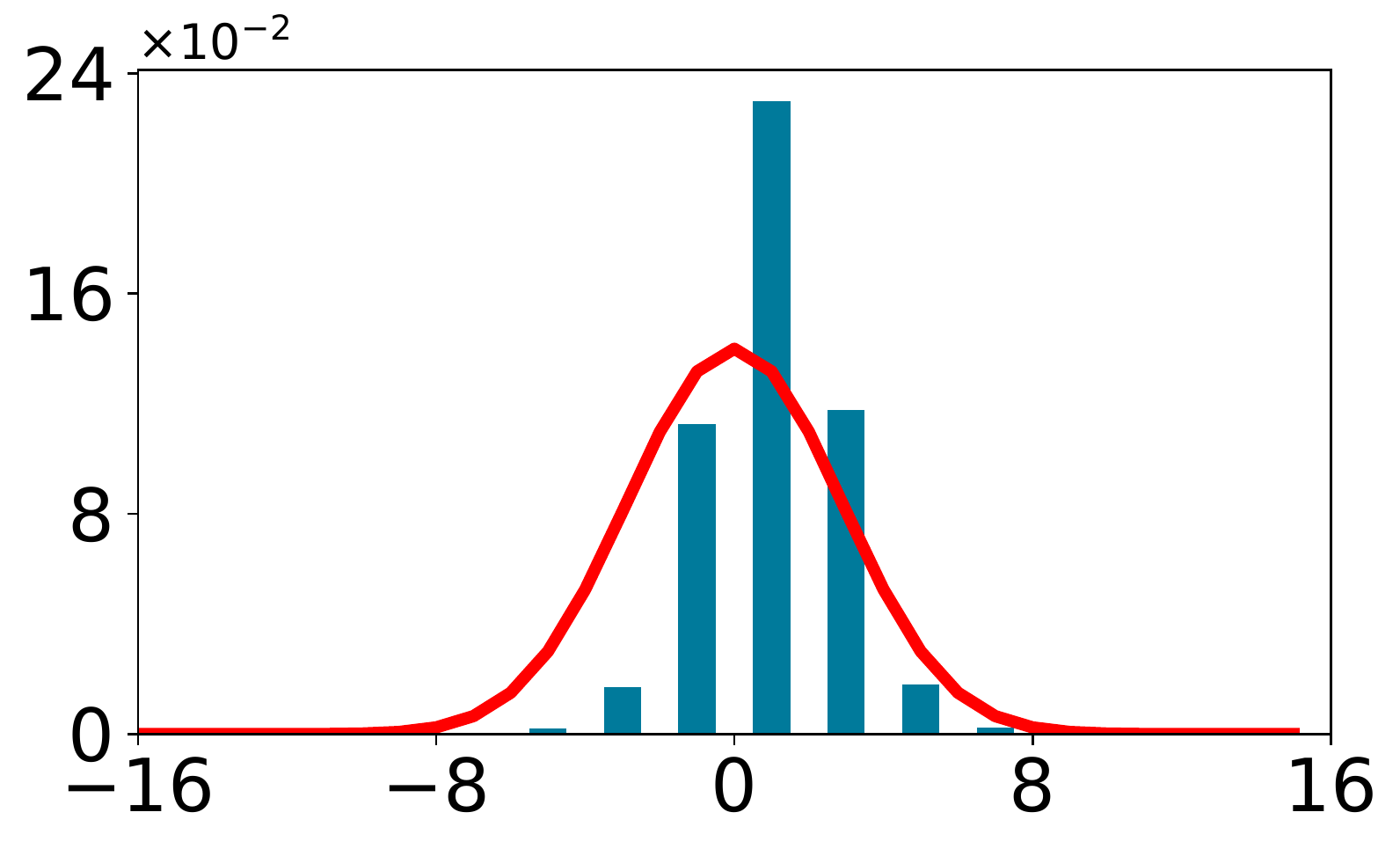}%
		}
		\subcaptionbox{Bar\label{sf:bar-rand}}{%
		\centering
			\includegraphics[width=0.42\textwidth]{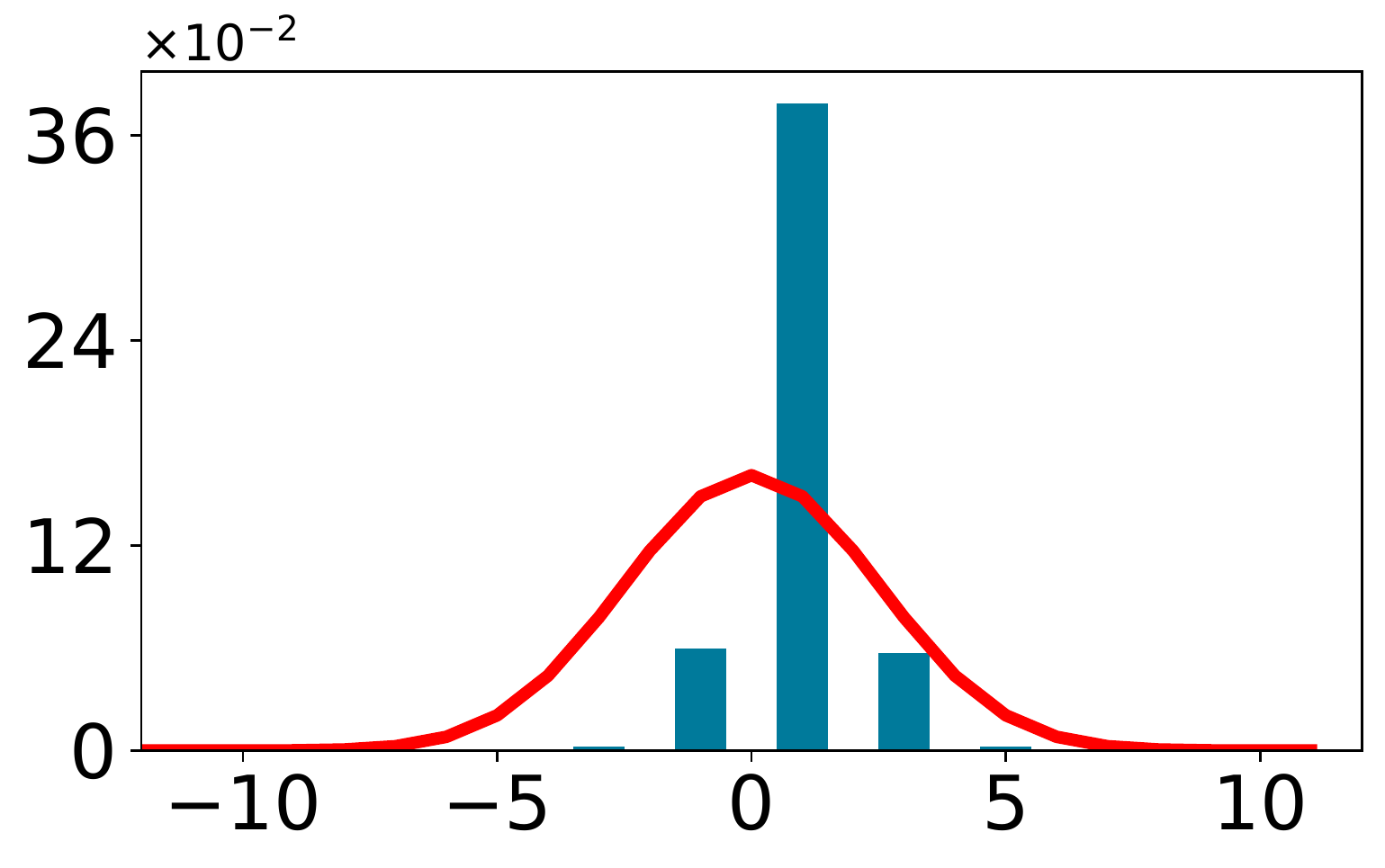}%
		}
		\caption{Distribution of fingerprint random walks for different sensors. Expected binomial distribution in red.}
		\label{fig:res-randomness}
  \end{minipage}
\end{figure}

Above, we have shown how much reduction in pairing time can be achieved by using \gls{fpake} instead of fuzzy commitments. 
\name also utilizes sensor fusion as compared to a single sensor modality used by state-of-the-art \gls{zip} schemes. 
The effect of sensor fusion on pairing time can be seen in~\autoref{tab:pairing-time} and~\autoref{fig:pair-time-data}. 
The maximum reduction of pairing time is proportional to the number of used sensors (e.g., three times shorter with three sensors as compared to one), assuming each sensor obtains the same number of bits in the same amount of time. 
From~\autoref{tab:pairing-time} (cf. \gls{fpake} pairing time), we see that for \textit{Acv}, \textit{Ach}, and \textit{Gyr} (each requires 10 seconds to produce one fingerprint) the reduction in pairing time is 2--3 times when fusing all three sensors. 
It is smaller in some cases than the maximum possible reduction because each sensor outputs a different number of bits after error correction due to varying similarity thresholds. 
In our setting, sensor fusion allows \name to shorten pairing time by an extra 2--3 times in addition to what is gained by \gls{fpake}. 
Combining more sensors would further reduce pairing time.

\subsection{Entropy of Fingerprints}
\label{sub:eval-fp-rand}
To evaluate entropy of fingerprints produced by \name, we (1) examine them for biases (e.g., bit patterns) and (2) estimate their min-entropy. 
To identify biases, we represent our fingerprints as random walks, with 1- and 0-bits showing steps in positive and negative directions~\cite{Bruesch:2019, Fomichev:2019perils}. 
The result follows a binomial distribution if fingerprints are uniformly random. 
We also study bit transition probabilities, interpreting each bit position in a fingerprint as a state in a Markov chain.
\autoref{fig:res-randomness} depicts the results of random walks for individual sensors.
The distributions for all sensors are centered around the mean, indicating that overall fingerprints have the equal number of 0- and 1-bits.
We see that more unique fingerprints can be generated from modalities with higher variation (e.g., \textit{Acv}). 
The Markov property is close to 0.5 for all sensors, showing that the probability of each bit in a fingerprint to be 0 or 1 is equal.  
These findings reveal no biases in our fingerprints, indicating that our quantization achieves its design goals (cf.~\autoref{sec:sysdesign}). 

\begin{figure}
\centering
  \includegraphics[width=\linewidth]{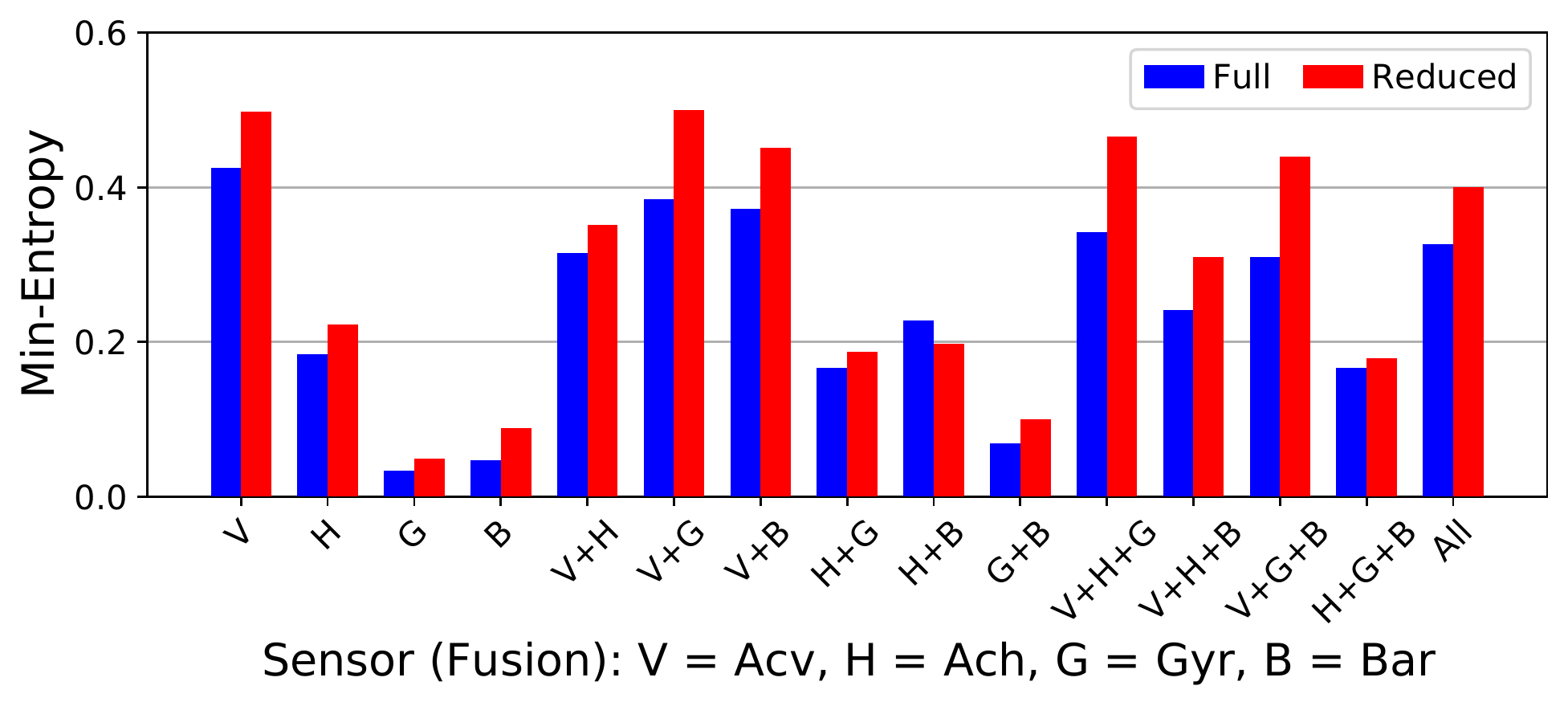}
\caption{Min-entropy of \name fingerprints estimated by NIST SP800-90B test suite (entropy in 1 bit).}
\label{fig:nist-800-90b-entropy}
\end{figure}

To assess min-entropy, we apply the NIST \textit{SP 800-90B} test suite~\cite{Turan:2018, NIST:2019}.
It consists of ten entropy estimators and is widely used~\cite{Zenger:2016, Kreiser:2018, Camara:2019}.
\autoref{fig:nist-800-90b-entropy} shows the estimated min-entropy for fingerprints of individual and fused sensors.
We obtain 0.43 bits of entropy for \textit{Acv}, 0.19 bits for \textit{Ach}, and below 0.05 bits for both \textit{Gyr} and \textit{Bar}, confirming our findings in~\autoref{fig:res-randomness}. 
Sensor fusion has a positive impact on min-entropy, which either stays close to the highest min-entropy in the combination or exceeds it. 
The fact that min-entropy increases when combining different sensors, indicates that they are uncorrelated, preventing the adversary from inferring one sensor signal from another.  
We consider the obtained entropy results to be conservative because the SP 800-90B suite is known to underestimate min-entropy~\cite{Zhu:2017}, and it makes a fair assessment given \textgreater $10^{6}$ data samples, which we do not have.
We find that dependency between consecutive bits is a decisive factor in lowering min-entropy of our fingerprints.
To check if this is caused by quantization parameters, we halve the number of bits in our fingerprints (cf. \textit{Reduced} in~\autoref{fig:nist-800-90b-entropy}), seeing only a modest increase in min-entropy. 
Thus, min-entropy in our fingerprints is restricted by the lack of entropy in the sensor data. 
In~\autoref{sec:discuss}, we elaborate on attainable entropy from our sensor data.

\autoref{fig:nist-800-90b-entropy} shows that the majority of fused fingerprints have 30--40\% of truly random bits. Thus, we need to collect more data to provide security, increasing pairing time of \name by 2.5--3 times to 75--120 seconds. 
For state-of-the-art \gls{zip} schemes, pairing time grows by 7 times, reaching several minutes, as they are more affected by non-random bits due to longer fingerprints. 

\subsection{Prototype Performance}
\label{sub:eval-prot}
\begin{figure}
\centering
  \includegraphics[width=\linewidth]{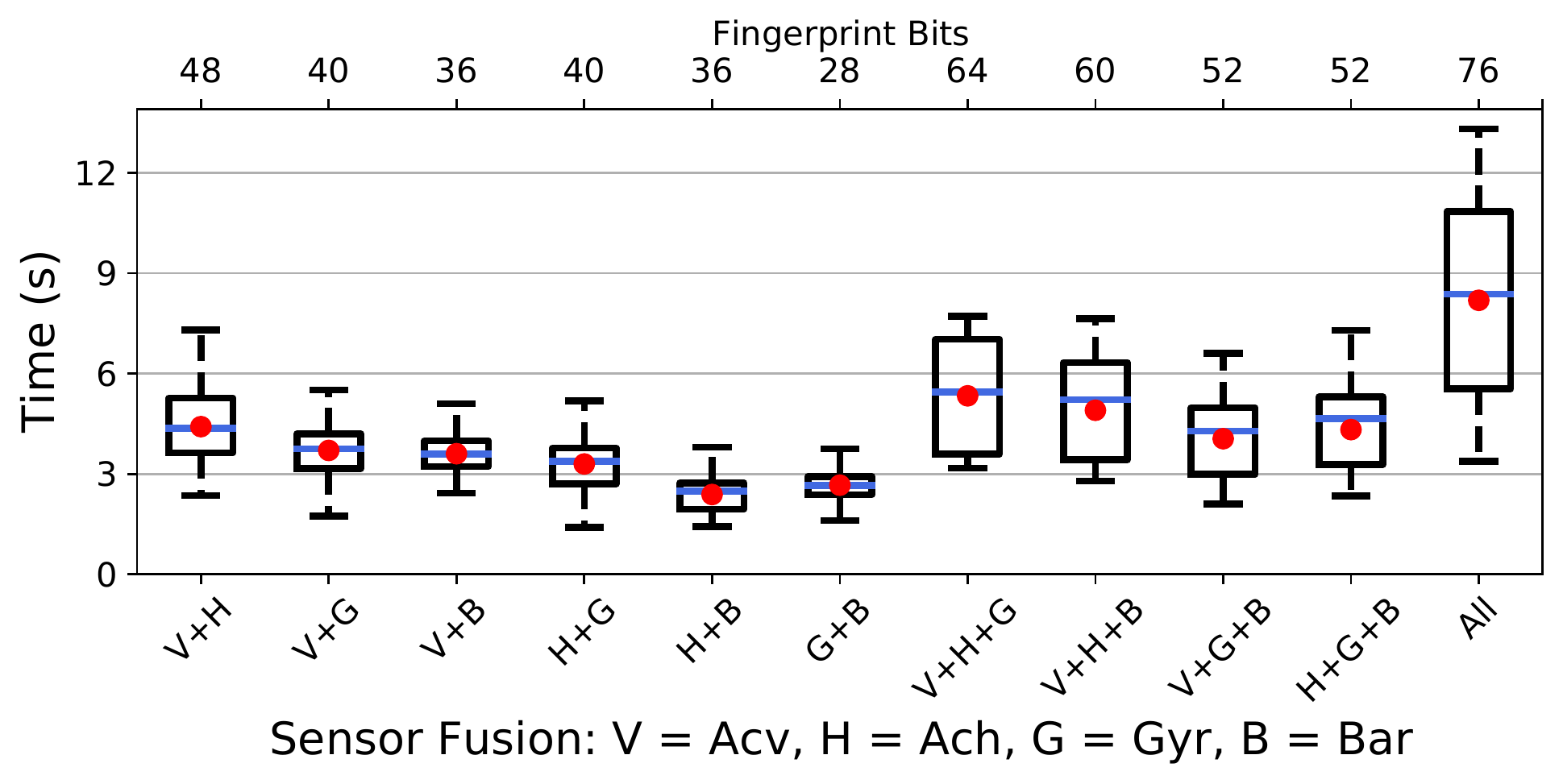}
\caption{Performance of \name for 128-bit key output.}
\label{fig:pairing-time-prototype-calculation-128bit}
\end{figure}

We benchmark our \name prototype on the Raspberry Pi 3 Model B, recording its performance in terms of computation and communication overhead. 
Specifically, we randomly sample 2000 fingerprints for each fusion combination, deploying them on two Raspberry Pis (i.e., 1000 fingerprints on each) connected via a Wi-Fi router. 
We measure the execution time to establish a 128-bit symmetric key on each device, showing the average performance in~\autoref{fig:pairing-time-prototype-calculation-128bit}\footnote{We use fingerprints from our evaluation. Accounting for  entropy loss in the fingerprints (cf.~\autoref{tab:pairing-time}) will increase the execution time by a few seconds.}.  
We observe a maximum time of around 4.4 seconds for two sensors (i.e., \textit{Acv + Ach}), growing to 8.2 seconds when fusing all of them. 
The execution time depends on the fingerprint size, and its deviation from the average performance increases with a lower similarity threshold (e.g., \textit{Acv + Ach} vs. \textit{Gyr + Bar}). 
We see that 60--80\% of the execution time accounts for the communication overhead, which can be reduced using a direct link between devices. 
Rising the output key size from 128 to 244 bits proportionally increases the execution time. 
Overall, \name runs efficiently on off-the-shelf \gls{iot} devices, imposing only a few seconds of overhead.   
Our prototype is Python-based without performance optimization techniques. Its building blocks (cf.~\autoref{subsub:prot}) can be reimplemented in C to deploy \name on more constrained devices. 

\section{Discussion}
\label{sec:discuss}
We provide relevant discussion points for \name.
\\
\textbf{Generalizability.} 
We show how to adapt the building blocks of \name: activity filter, quantization, and \gls{fpake} to be used for \gls{zip} in other use cases (e.g., smart home, wearables). 
Our activity filter utilizes generic metrics: average power, \gls{snr}, and number of prominent peaks that can be computed on any sensor signal.
To find metrics thresholds, excluding low-entropy signals, we examine metrics of sensor signals of fixed length (e.g., 10 seconds), capturing strong and weak ambient activity. 
With this approach, we obtain thresholds suitable for different cars and road types.
Similar results are reported for the average power threshold of audio signals recorded in different places~\cite{Karapanos:2015}. 
Hence, thresholds for activity filter metrics can be determined once per use case and sensor type.
Our activity filter can be easily adapted by wearable \gls{zip} utilizing human gait captured by the acceleromter.
Specifically, all three metrics (prominent peaks can mark gait cycles) are computed on gait signals of chosen length (e.g., 30 seconds), while metrics thresholds can be derived using public gait data of moving and still users~~\cite{Schurmann:2017}.

\name quantization has worked well on four sensor modalities. 
To apply it for other \gls{zip} use cases, two parameters need to be adjusted: (1) length of input sensor signal and (2) number of output fingerprint bits.  
These parameters are set empirically based on the duration and variation of scenario-specific ambient activity captured by the sensor signal. 
For example, in smart home, a door knock event lasting a few seconds can be recorded by the microphone and accelerometer~\cite{Han:2018}. 
The former signal has higher variation, thus our quantization can be set to output more fingerprint bits from it.

Our \gls{fpake} findings (cf.~\autoref{tab:fingerprintsizes}) are generic for a given similarity threshold and security level, hence directly reusable by other \gls{zip} schemes. 
For a different choice of similar threshold/security level, the required fingerprint size in bits providing protection against offline attacks can be computed, as explained in~\autoref{sec:sysdesign}.
\\
\textbf{Entropy of Sensor Data.}
The min-entropy presented in~\autoref{fig:nist-800-90b-entropy} results from sensor data collected on high-quality flat roads, giving the lower bound of attainable entropy.
We do not cover gravel, forest, or mountain roads that have profound bumpiness (\textit{Acv}) and sharp turns (\textit{Gyr}). 
Also, we do not have representative data from hectic metropolis driving, which should reveal distinct acceleration patterns (\textit{Ach}) as well as from hilly regions with rapidly changing altitude (\textit{Bar}). 
These different road and traffic conditions have high potential for increasing entropy in sensor data~\cite{Han:2017}. 
Another way of obtaining more entropy from sensor data is customized quantization. Our quantization focuses on (1) extracting bits from heterogeneous sensors and (2) reducing entropy biases in fingerprints, hence it may not be optimal in the amount of attainable entropy. 
Prior work explores various quantization methods~\cite{Bruesch:2019, Groza:2020} some of which can be adapted to \name.
\\
\textbf{Deployment Considerations.}
To deploy \name in a real car setting a few points need to be considered. 
First, devices are expected to continuously sense their context before they establish pairing, eliminating the need for time synchronization~\cite{Han:2018, Miettinen:2018} (i.e., each device extracts fingerprints bits from common parts of context passing the activity filter).
In other words, devices observe common context events (e.g., road bump) in the same timeline (i.e., similar to~\cite{Han:2018}) and can buffer them, tolerating clock offset between devices. 
For this to work, devices must maintain the same sampling rate of context measurements and start them simultaneously (e.g., upon a broadcasted command). 
For example, a major component of the car (e.g., infotainment system) can broadcast such a command when a car is started. 
Since devices in the same car are located nearby, they will receive this command almost at the same time. 
To further eliminate the effect of different devices receiving the broadcasted command at negligibly different times and account for overhead to trigger sensing, devices can start measuring context upon the command reception after a short pause (e.g., 5 seconds); for this they do not need synchronized clocks as well. 
Furthermore, \name extracts much fewer bits from context signals (e.g., 24 bits from 10 seconds) as compared to existing \gls{zip} schemes (e.g., 128 bits from 5 seconds in~\cite{Xu:2016} or 512 bits from 6 seconds in~\cite{Schurmann:2013}), making \name less susceptible to several millisecond offsets between these signals. 
Specifically, we try injecting 5--7 millisecond offsets between context signals that we used to evaluate \name, finding that it would reduce \glspl{tar} of individual sensors (cf.~\autoref{sf:tar-indiv}) by maximum 10\% for \textit{Acv}, 7\% for \textit{Ach}, and below 5\% for \textit{Gyr} and \textit{Bar}, while \glspl{far} remain the same. 
We consider this reduction to be acceptable for a proof-of-concept \name, however, further research can investigate how to eliminate the effect of synchronization errors in real deployments.
Since \name requires a few dozen seconds to pair, collecting context for this time using low-power sensors will not impose much overhead.  
Second, each device is expected to learn parameters of the scheme (e.g., quantization) prior to pairing: in \name it can happen upon the scheme installation, as commonly assumed in \gls{zip}~\cite{Han:2018, Schurmann:2017, Wu:2020}.
Before trying to pair, each device can advertise its desired security level (e.g., 128- or 244-bits in \gls{fpake}), pairing with those devices that support the same security level.  
\textbf{Limitations.}
We evaluate \name using devices fixed inside a car interior, covering the likely use case of pairing between a mounted user device (e.g., smartphone) and an infotainment system.
However, users may interact with their devices, affecting accelerometer and gyroscope readings.
Differentiating between human and vehicle motion in the sensor data collected inside a moving car is an open research question~\cite{Chen:2017}.
We envision that predicting sensor data resulted from human motion~\cite{Wu:2020} and filtering it afterwards~\cite{Schurmann:2017} can help address this question.

\section{Related Work}
\label{sec:rwork}


To date, a number of \gls{zip} schemes utilizing various sensors (e.g., microphone, accelerometer) to capture context have been proposed~\cite{Schurmann:2013, Miettinen:2014, Schurmann:2017, Han:2017, Han:2018, Miettinen:2018, Lin:2019}.
The state-of-the-art \gls{zip} schemes rely on the fuzzy commitments cryptographic primitive~\cite{fuzzyCom} to establish a shared secret key. Other cryptographic alternatives include customized extensions of fuzzy commitments~\cite{Schurmann:2017} or the \gls{eke} protocol~\cite{Groza:2020}. 
However, these extensions do not have proven security guarantees.
The majority of proposed \gls{zip} schemes rely on a single common sensor to capture context. 
The existing schemes utilizing fuzzy commitments and context based on a single sensor modality suffer from (1) prolonged pairing time, (2) vulnerability to offline attacks, and (3) attacks caused by the predictable context (e.g., replay).   
\name overcomes these limitations by a novel design, namely combining the \gls{fpake} protocol~\cite{Fpake:2018} and multi-sensor context constructed by combining multiple sensor modalities (i.e., sensor fusion).

\begin{table}
\small
\centering
	\caption{Comparison with state-of-the-art \gls{zip} schemes.}
	\label{tab:zip-cmp}
  \begin{tabular}{c|cccc}
  	\toprule
  	Scheme & Use Case  & Time (s) & (FAR, FRR) & Bias \\
  	\midrule
  	Sch{\"u}rm. \& Sigg~\cite{Schurmann:2013}{\large $^\dagger$} & In-car & 120 & (0.10, 0.10) & Low \\
  	Miettinen et al.~\cite{Miettinen:2014}{\large $^\dagger$} & In-car & 1280 & (0.23, 0.23) & High \\
  	Convoy~\cite{Han:2017} & In-car & 300 & - & - \\
  	Miettinen et al.~\cite{Miettinen:2018} & Home & 5640 & (0.03, 0.02) & - \\
  	Perceptio~\cite{Han:2018} & Home  & 8280 & - & - \\
  	BANDANA~\cite{Schurmann:2017} & Wearables  & 96 & - & High \\
  	\hline
  	\name & In-car  & 20 & (0.0, 0.06) & Low \\
  	\bottomrule
  \end{tabular}
      \smallskip\centering
  \center{{\large $^\dagger$}evaluated in~\cite{Fomichev:2019perils}. We show best achievable results for each scheme.}
  \label{tab:comparison}
\end{table}

\autoref{tab:zip-cmp} compares \name and prominent state-of-the-art \gls{zip} schemes in terms of pairing time, error rates, and entropy biases in the fingerprints. 
We note that this comparison is indicative, as we use the information reported in the original publication for each \gls{zip} scheme.
\name has the shortest pairing time among the schemes, including those that are used for in-car pairing, while achieving low error rates. 
This shortest pairing time is due to the combination of \gls{fpake} and sensor fusion, which can together give a 3--9 reduction in pairing time (cf.~\autoref{sub:eval-pair-time}). 
However, pairing time also highly depends on the used context (e.g., continuous gait~\cite{Schurmann:2017} vs. infrequent knock~\cite{Han:2018}) and quantization method (e.g., in~\cite{Miettinen:2014} one bit is derived from two minutes of sensor data.)

The schemes~\cite{Schurmann:2013} and~\cite{Miettinen:2014} utilizing ambient audio and noise levels, respectively, are evaluated for in-car pairing~\cite{Fomichev:2019perils}, showing error rates above 0.1. 
Despite audio and noise level context varying significantly in a running car, the fingerprints of those schemes contain entropy biases (e.g., more 0-bits). 
\textit{Convoy} that uses road bumpiness captured by the accelerometer for pairing is vulnerable to the context replay attack~\cite{Han:2017}, however the resulting \gls{far} is not reported. 
A similar work bears the same weakness as \textit{Convoy} but does not state the pairing time~\cite{kim2019secure}. 
\gls{zip} schemes for pairing smart home devices~\cite{Han:2018, Miettinen:2018} may achieve comparable error rates to \name, requiring, however, at least two orders of magnitude longer time.  
This time will further increase in the case of entropy biases, which are not evaluated by the considered schemes. 
We note that the longest pairing time of \textit{Perceptio}~\cite{Han:2018} is a tradeoff, as the scheme enables pairing between devices with heterogeneous sensors (e.g., microphone and accelerometer). 
For \gls{zip} schemes targeting wearables such as BANDANA~\cite{Schurmann:2017}, utilizing human gait captured by the accelerometer, the pairing time is closest to \name. 
However, such schemes often show  bit patterns in their fingerprints and are vulnerable to video-based attacks~\cite{Bruesch:2019}.

Our review of related \gls{zip} work reveals important results: entropy biases of various level of severity exist in fingerprints of all schemes.  
This is worrying, as the state of the art relies on fuzzy commitments, where high entropy of fingerprints is imperative to prevent offline attacks. 
Also, none of the works explicitly accounts for entropy biases (e.g., by saying how many more bits need to be collected). 
The impact of entropy biases is less severe in \gls{fpake}, as it limits the offline attack in time and number of attempts. 
We notice that many \gls{zip} schemes use previous versions of NIST statistical tests~\cite{rukhin2001statistical} to find entropy biases, reporting results for only passed tests, without further investigation~\cite{Xu:2016, Lee:2019, Lin:2019}.  
Thus, we urge researchers to scrutinize the entropy of fingerprints derived from context with recent NIST tests~\cite{Turan:2018} and additional tools such as in~\cite{Bruesch:2019, Fomichev:2019perils}.


\glsresetall

\section{Conclusion}
\label{sec:concl}
In the age of the \gls{iot} securing wireless communication of smart devices is crucial to protect their data. 
\Gls{zip} allows establishing a shared secret key between devices based on their physical context (e.g., ambient audio).
We propose \name, a novel \gls{zip} scheme that significantly reduces pairing time, while providing stronger security than state-of-the-art \gls{zip} schemes. 
The main contribution of \name is its innovative design combining the \gls{fpake} protocol and sensor fusion.  
We implement and empirically evaluate \name in the exemplary use case of intra-car device pairing, demonstrating that \name (1) reliably pairs devices inside the same car, achieving up to three times faster pairing than state-of-the-art \gls{zip} schemes, (2) is secure against various attacks, and (3) runs efficiently on off-the-shelf \gls{iot} devices.


\section*{Acknowledgments}
\label{sec:ack}
We thank our shepherd and our anonymous reviewers for their insightful comments that helped to improve this paper.
We also thank Max Maass and Arne Br{\"u}sch for their assistance in conducting this research.  
This work has been co-funded by the Research Council of Norway as part of the project Parrot (311197) as well as the German Federal Ministry of Education and Research and the Hessian Ministry of Higher Education, Research, Science and the Arts within their joint support of the National Research Center for Applied Cybersecurity ATHENE. 
Jun Han and Julia Hesse are co-corresponding authors of this work.


\bibliographystyle{abbrv}
\bibliography{bibliography}  

\end{document}